\documentclass[manuscript]{acmart}

\usepackage{xspace}   
\usepackage{textcomp} 
\usepackage{graphicx} 
\usepackage{xcolor} 
\usepackage{multirow}
\usepackage{booktabs}
\usepackage{array}

\makeatletter
\@ifclasswith{acmart}{revise}{

  \newcommand{\revise}[1]{\textcolor{blue}{#1}}
}{

  \newcommand{\revise}[1]{#1}
}
\makeatother

\DeclareRobustCommand{\projectname}{\textsc{CareMap}\xspace}

\AtBeginDocument{%
  }

\setcopyright{acmlicensed}
\copyrightyear{2018}
\acmYear{2018}
\acmDOI{XXXXXXX.XXXXXXX}

\acmConference[Conference acronym 'XX]{Make sure to enter the correct
  conference title from your rights confirmation emai}{June 03--05,
  2018}{Woodstock, NY}
\acmISBN{978-1-4503-XXXX-X/18/06}

\begin{document}

\title[Information Needs and Design Opportunities for Family Caregivers of Older Adult Patients in Critical Care]{``It Felt Like I Was Left in the Dark'': Exploring Information Needs and Design Opportunities for Family Caregivers of Older Adult Patients in Critical Care Settings}


\author{Shihan Fu}
\affiliation{%
  \institution{Northeastern University}
  \city{Boston}
  \state{Massachusetts}
  \country{United States}}
\email{fu.shiha@northeastern.edu}

\author{Bingsheng Yao}
\affiliation{%
  \institution{Northeastern University}
  \city{Boston}
  \state{Massachusetts}
  \country{United States}}
\email{b.yao@northeastern.edu}

\author{Smit Desai}
\affiliation{%
  \institution{Northeastern University}
  \city{Boston}
  \state{Massachusetts}
  \country{United States}}
\email{sm.desai@northeastern.edu}

\author{Yuqi Hu}
\affiliation{%
  \institution{Northeastern University}
  \city{Boston}
  \state{Massachusetts}
  \country{United States}}
\email{hu.yuqi@northeastern.edu}

\author{Yuling Sun}
\affiliation{%
  \institution{University of Michigan}
  \city{Ann Arbor}
  \state{Michigan}
  \country{United States}}
\email{}

\author{Samantha Stonbraker}
\affiliation{%
  \institution{University of Colorado College of Nursing}
  \city{Aurora}
  \state{Colorado}
  \country{United States}}
\email{Samantha.Stonbraker@CUAnschutz.edu}

\author{Yanjun Gao}
\affiliation{%
  \institution{University of Colorado Anschutz Medical Campus}
  \city{Aurora}
  \state{Colorado}
  \country{United States}}
\email{}

\author{Elizabeth M. Goldberg}
\affiliation{%
  \institution{University of Colorado Anschutz Medical Campus}
  \city{Aurora}
  \state{Colorado}
  \country{United States}}
\email{elizabeth.goldberg@cuanschutz.edu}

\author{Dakuo Wang}
\authornote{Corresponding author}
\affiliation{%
  \institution{Northeastern University}
  \city{Boston}
  \state{Massachusetts}
  \country{United States}}
\email{d.wang@northeastern.edu}

\renewcommand{\shortauthors}{Fu et al.}

\begin{abstract}
Older adult patients constitute a rapidly growing subgroup of Intensive Care Unit (ICU) patients. 
In these situations, their family caregivers are expected to represent the unconscious patients to access and interpret patients' medical information. 
However, caregivers currently have to rely on overloaded clinicians for information updates and typically lack the health literacy to understand complex medical information.
Our project aims to explore the information needs of caregivers of ICU older adult patients, from which we can propose design opportunities to guide future AI systems. 
The project begins with formative interviews with 15 caregivers to identify their challenges in accessing and understanding medical information; 
From these findings, we then synthesize design goals and propose an AI system prototype to cope with caregivers' challenges. 
The system prototype has two key features: a timeline visualization to show the AI extracted and summarized older adult patients' key medical events; and an LLM-based chatbot to provide context-aware informational support.
We conclude our paper by reporting on the follow-up user evaluation of the system and discussing future AI-based systems for ICU caregivers of older adults.
\end{abstract}

\ccsdesc[500]{Human-centered computing~Human computer interaction (HCI)}
\ccsdesc[500]{Human-centered computing~HCI design and evaluation methods}
\ccsdesc[300]{Human-centered computing~Collaborative and social computing theory, concepts and paradigms}
\ccsdesc[500]{Applied computing~Life and medical sciences}

\keywords{Human-AI collaborations, family caregivers, critical care, large language models, information need}

\received{20 February 2007}
\received[revised]{12 March 2009}
\received[accepted]{5 June 2009}

\begin{teaserfigure}
    \centering
    \includegraphics[width=1\linewidth]{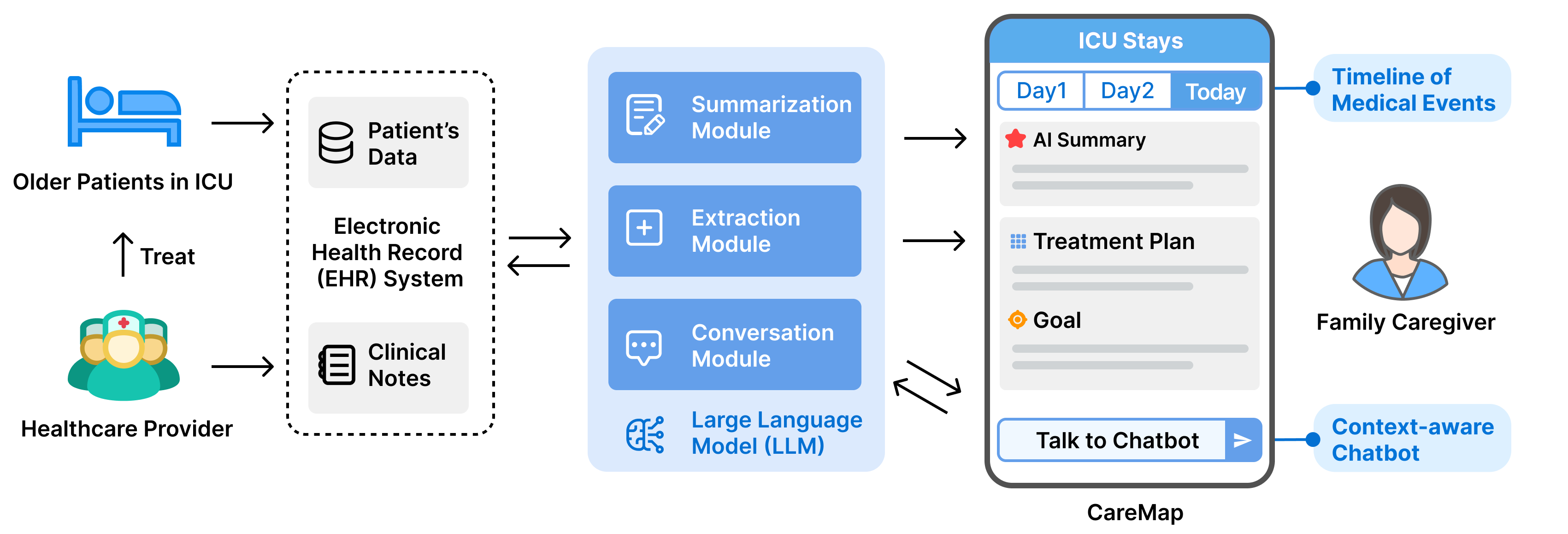}
    \caption{The structure of the CareMap system. CareMap takes raw EHR data, such as patient information and clinical notes, as input and transforms it into caregiver-facing outputs. These outputs are presented through two key features: a visual timeline of major medical events and a chatbot that supports real-time, personalized inquiries. The system is powered by three LLM-driven modules: summarization, information extraction, and conversational support. \projectname is designed to enhance caregivers’ access to and understanding of a patient’s clinical course in critical care.}
    \label{teaser}
\end{teaserfigure}

\maketitle

\section{Introduction}

The Intensive Care Unit (ICU) is a specialized critical care ward in hospitals, which is dedicated to providing acute, life-saving treatment for patients with life-threatening conditions caused by severe illnesses~\cite{hanberger2005intensive, valentin2011recommendations}. 
In the United States, older adults (aged 65 and above)
account for more than half of ICU admissions, with the amount continuing to grow~\cite{angus2006critical, flaatten2017status, jarvis2023physical}.
Such a vulnerable population of ICU patients has been the focus of care provision~\cite{goldfarb2017outcomes, grieve2019analysis}, however, caregivers of older adult patients are often overlooked in such critical care settings~\cite{gonccalves2023surviving}.
Caregivers of older adult patients are often their family members and are assigned powers of attorney to unconscious patients for making critical medical decisions~\cite{au2017family, hines2011effectiveness, johnson1995perceived}. 
Prior research suggests that if the family caregivers can actively engage with the medical decision-making process, it can significantly enhance the ICU patients' final outcomes~\cite{choi2019exploring}.

However, caregivers often find it challenging to access and understand the necessary clinical information to make informed decisions on behalf of their care recipients in the ICU~\cite{gaeeni2014informational, al2013family, paul2004meeting}. 
Firstly, caregivers can only access limited information about the ICU patients, as they cannot enter the ICU ward frequently, their older adult patients are often unable to talk to them due to unconsciousness, and they primarily rely on the ICU clinical team, who are already overwhelmingly busy, to obtain critical information.~\cite{schnock2017identifying, au2017family, hines2011effectiveness}. 
Secondly, caregivers often find it difficult to fully understand the highly specialized medical language or results due to their low level of health  literacy~\cite{young2017family}.
To address these challenges, various solutions have been proposed, yet mostly centered on the ICU clinical team side to innovate their best practices. 
For example, ICU clinicians are recommended to communicate more frequently with caregivers~\cite{pronovost2003improving}, and they
may use a checklist~\cite{caligtan2012bedside, nelson2006improving} or a daily-goal form~\cite{scheunemann2011randomized, azoulay2002impact}to present more complete clinical information and in a more understandable way to caregivers.
However, these best-practice recommendations often fail to meet the caregivers' information needs~\cite{davidson2017guidelines, goldfarb2017outcomes, meert2013family}.
To truly address this gap, we need novel, caregiver-centered perspectives and technical solutions that directly support their access to and understanding of ICU-related information.

Researchers in Human-Computer Interaction (HCI) and Computer-Supported Cooperative Work (CSCW) have proposed various technological solutions to support caregivers' information needs~\cite{Sette2023, Yuexing2024, Nikkhah2021, Eunkyung2024}.
One group of works designed new tools to facilitate clinicians sharing information with caregivers~\cite{ziqi2024, Yuexing2024} and caregivers seeking clinical information of older adult patients~\cite{montagna2023data, wei2024leveraging}.
Another group of works focused on the development of visualization tools to decipher complex clinical information to support caregivers' understanding (e.g., a visualization dashboard for caregivers to understand autism children's behaviors~\cite{kong2017comparative} and a dashboard for caregivers to support patients' upper-limb stroke rehabilitation therapy process~\cite{ploderer2016armsleeve}).
However, these designs remain limited in the ICU context: as these tools are not tailored to support caregivers' information needs in high-stakes and time-sensitive ICU settings; and the complexity and density of information generated by the ICU environment can easily overwhelm caregivers in comparison to the other contexts. As such, existing visualization dashboards may be insufficient to support caregivers in effectively understanding patient information in critical care environments.

Recent advancements in artificial intelligence (AI) have opened new possibilities to help family caregivers better access and easily understand clinical information. 
For instance, AI models demonstrate great potential to extract relevant information from massive and unrelated raw clinical data from electronic health records (EHRs)~\cite{hayrinen2008definition}, and AI can also help translate complicated medical jargon into simplified language for the general audience to understand ~\cite{yim2024preliminary, wachter2024will, wong2018using}. 
In addition, chatbot systems powered by large language models (LLMs) can support caregivers' individualized information-seeking requests in real time~\cite{afshar2024prompt}. 
However, it remains underexplored whether and how these AI-based technologies may effectively support caregivers' information needs while their older adult patients are in the ICU.
To this end, we propose this study to focus on exploring the information needs of caregivers of ICU older adult patients and designing novel AI functionalities to support their needs within the critical care environment. 

Specifically, we conducted a two-stage study.
First, we conducted a formative study consisting of semi-structured interviews with 15 caregivers to understand the challenges they faced in how they access and understand clinical information in critical care settings.
Caregivers reported that the primary difficulties come from fragmented clinical information updates from healthcare providers, complicated clinical reports with limited or no explanations available, and frustration caused by repetitive but ineffective communication with the clinical team.

Building on these findings, we propose a set of design guidelines and implement an AI-based prototype system, \projectname, to support caregivers’ access to and understanding of ICU patient information. As illustrated in Figure~\ref{teaser}, the system processes complex clinical information from the ICU and leverages three LLM-driven modules—summarization, extraction, and conversation—to generate caregiver-facing outputs, including daily summaries, treatment plans, and context-aware responses. 
These outputs are delivered through two core features: (1) a visual timeline of key medical events that streamlines access to systematic status updates, and (2) an LLM-based chatbot that enables personalized information inquiries through natural language conversation. \revise{We then conducted usability evaluations with 10 family caregivers to assess the prototype’s usability, and with 2 clinical experts to evaluate the quality of the system-generated outputs as well as provide usability feedback from a clinical perspective.}
Qualitative analysis revealed the potential of LLM-based systems to reduce caregivers’ cognitive burdens, improve their access to complex clinical information, and enhance engagement with care teams.
 

This work makes the following contributions. 
First, we systematically investigated the challenges caregivers of older adults encountered in accessing and understanding clinical information in critical care settings.
Then, we presented a suite of design guidelines to support the design of caregiver-facing interfaces with advanced AI technologies and evaluated two AI-based functionality designs.
Finally, our findings contribute to the broader discourse on technology-mediated caregiver support and the role of AI in supporting caregiving experiences for family caregivers in high-stakes, time-sensitive critical care scenarios.

\section{Related Work}~\label{relatedwork}

We conduct a comprehensive review of the experiences of family caregivers for older adult patients in critical care environments, various technologies developed for caregivers in clinical settings, and the role of AI-based clinical information technology in this context.
\subsection{Family Caregivers of Older Adults in Critical Care Settings}

In the United States, over half of ICU admissions involve older adults aged 65 and above~\cite{flaatten2017status}. 
This population is receiving growing attention due to their complex health needs, which are often compounded by chronic diseases and acute health crises~\cite{jarvis2023physical}. 
When older adult patients are unable to talk or are unconscious due to severe injuries, family caregivers of older adults have to actively engage in the care process~\cite{au2017family, johnson1995perceived}.
For instance, family caregivers could be assigned powers of attorney to make critical decisions on behalf of their loved ones~\cite{hines2011effectiveness}.
On the other hand, they need to prepare for the patient’s post-discharge care by coordinating the necessary support with healthcare providers, as caregiving demands often continue or escalate following patients' ICU discharge~\cite{choi2014fatigue}. 
The demanding responsibilities of family caregivers often place an immense burden on them, which is often compounded by the difficulties of accessing and understanding medical information in the ICU.

However, caregivers are often not familiar with the critical care environment and only possess limited health literacy, which causes significant difficulty for them in understanding the overwhelming and complicated critical care situation.
ICU wards impose restrictive visitation policies to avoid disruptions caused by the presence of unnecessary non-professional personnel~\cite{dragoi2022visitation, tabah2022variation}. 
Such restrictions leave caregivers with difficulties in comprehensively accessing and understanding the patient's health condition and treatment progression in a timely manner.
Moreover, caregivers experience deep anxiety and helplessness about their loved ones because of the sudden deterioration of older adults' health and are unable to anticipate what could happen to the older adults~\cite{jennerich2020unplanned}.
Studies report that up to 54\% of family caregivers experience psychological trauma or severe stress after caregiving experiences in critical care situations~\cite{alfheim2019post}.


Consequently, caregivers have to rely heavily on direct communication with healthcare providers to stay informed about the patient’s status~\cite{yoo2020critical}. 
However, the provider teams are always overloaded with caring for critically ill patients and have very limited opportunities and time to talk with caregivers.
Such constrained caregiver-provider communications can hardly meet caregivers' information needs.
Research highlights persistent dissatisfaction among caregivers despite institutional efforts to improve clinical communication through tools like daily goals forms~\cite{pronovost2003improving}, critical care family satisfaction surveys~\cite{steel2008impact}, and allocation of additional resources to ICU wards~\cite{breslow2005technology}. 
Further, the lack of clear, timely, and actionable information heightens caregivers' stress and disrupts their ability to prepare for the patient’s post-discharge care~\cite{azoulay2016communication}.

\subsection{Technologies Supporting Family Caregivers in Clinical Settings}


In recent years, the fields of Computer-Supported Cooperative Work (CSCW) and Human-Computer Interaction (HCI) have emphasized the critical role caregivers play in patient care within clinical settings~\cite{Don2024, Siddiqui2023, Yunan2013} and emphasized the challenges caregivers face associated with clinical information~\cite{Stefanidi2023}. In response, prior research has explored various technological solutions to address these challenges~\cite{Pine2018}.

On the one hand, researchers have explored various technology designs to enhance clinical information accessibility, including facilitating the sharing of information between clinicians and caregivers~\cite{ziqi2024, Yuexing2024, Bowers2024} and assisting caregivers to seek clinical information of older adult patients~\cite{montagna2023data, wei2024leveraging}.
For example, ~\citet{Bowers2024} developed a mobile application that offers template support for caregivers to document and assess behavioral changes in patients with congenital heart disease, facilitating effective sharing of clinical information with clinicians. 
SaludConectaMX is a mobile application designed for caregivers to access clinical information, including patient medical history and oncology treatments, to assist in tracking patients' evolving health trajectories~\cite{Schnur2024}.
~\citet{Barbarossa2023} developed a dashboard to present the most relevant clinical information of older adult patients with dementia, including fall events, sleep data, and wellbeing data, for caregiver access to patient status. 
While effective, most of these tools are intended for caregivers of patients with chronic conditions, such as dementia.
Consequently, their accessibility is not suitable for the time-sensitive and high-stakes clinical environment of the ICU.

Another area of research explores using visualization tools to decipher complex clinical information to support caregivers' understanding of clinical information~\cite{kong2017comparative, Hong2017, Nyapathy2019, liu2011, Shea2019, Hwang2014}.
For instance,~\citet{kong2017comparative} introduced EnGaze, a web-based tool designed to visualize the communication behaviors of children with autism during clinical visits, aiming at improving caregivers' understanding of the condition and promoting their active involvement in care planning.
An interactive system designed by~\citet{Hong2017} assists families and clinicians in reviewing radiology imaging results by offering simplified definitions and diagrams of medical term during consultations.
~\citet{Nyapathy2019} presented a visualization mobile application designed to assist caregivers in recording and visualizing the long-term condition of asthma patients, thereby enhancing the shared understanding of the condition with clinicians.
While these technologies provide valuable assistance for caregivers in routine clinical settings, the complexity and density of information produced in the ICU can significantly overwhelm caregivers compared to other contexts.
The proposed visualization technology solutions are insufficient to support caregivers in understanding the clinical information of older adult patients in the ICU. 

In conclusion, existing tools and technologies remain inadequate in supporting caregivers of older adult ICU patients in accessing and understanding information.
Existing solutions are fragmented in scope, indicating a significant need to leverage novel technologies to systematically address caregivers' difficulties in accessing and understanding clinical information in critical care settings.

\subsection{AI-based Clinical Information Technologies}
 
Recent advancements in artificial intelligence (AI) have shown significant potential to improve caregivers' access to and understanding of clinical information. 

In order to improve caregivers' access to complex clinical information, researchers have developed AI models to extract complex clinical information from electronic health records (EHRs) through summarization~\cite{Sette2023, Nikkhah2021, gopinath2020fast}, identify high-risk factors in medical records~\cite{corey2018development}, and predict risks from clinical data, such as sepsis risk~\cite{suresh2017clinical, yin2024sepsiscalc}. 
AI can also enhance the understanding of clinical information.
For instance, some models can translate complicated medical jargon into accessible language for the general audience to understand ~\cite{yim2024preliminary, wachter2024will, wong2018using}, and simplify lengthy medical texts into shorter versions~\cite{basu2023med, artsi2024large}. In addition, chatbot systems based on large language models (LLMs) show potential in supporting caregivers' personalized information-seeking needs~\cite{afshar2024prompt}. 
Such solutions have proven effective in delivering health advice regarding screening, diagnosis, treatment, and disease prevention~\cite{huo2025large}. 
For example, \citet{ramjee2024cataractbot} developed an LLM-based chatbot named CataractBot that addresses patient inquiries regarding cataract surgery.

Overall, AI have demonstrated the potential to address the aforementioned challenges faced by caregivers of older adult patients in the ICU. 
However, there has been a lack of systematic investigation from the caregivers' perspective regarding their needs in accessing and understanding clinical information.
Moreover, how to design effective AI-based technologies to address caregivers' needs in accessing and understanding information remains underexplored.

\begin{figure}
    \centering
    \includegraphics[width=1\linewidth]{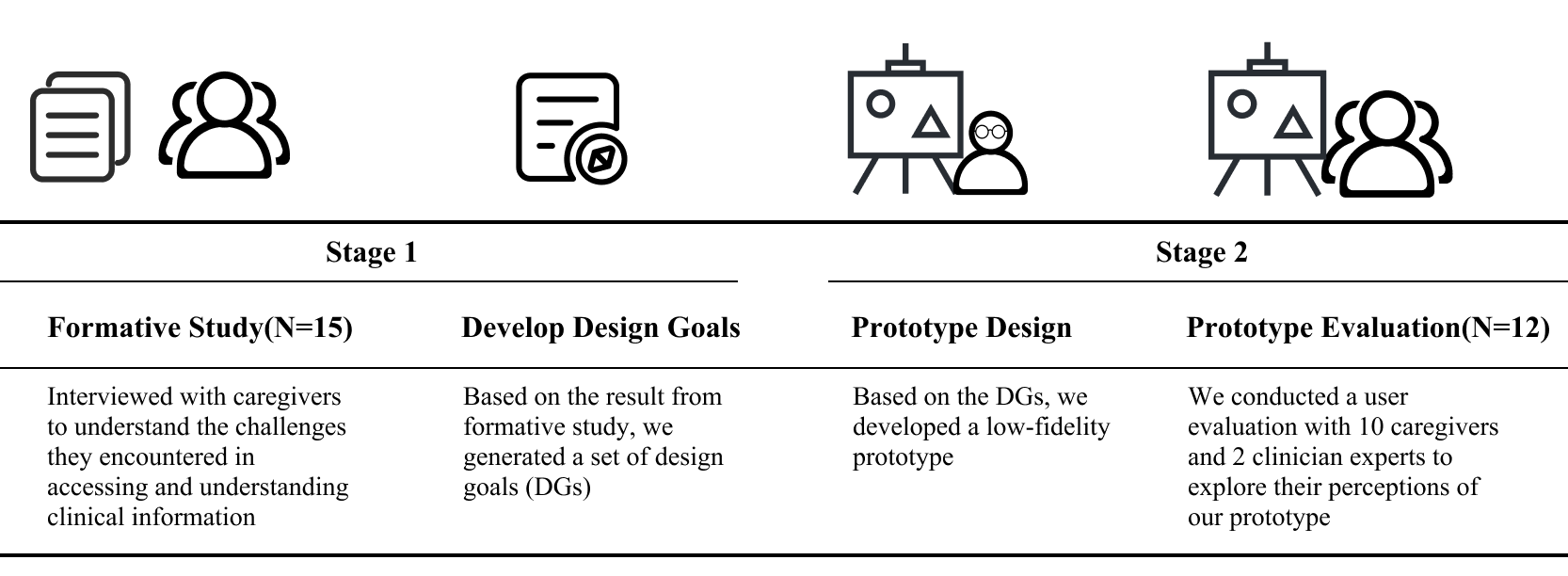}
    \caption{Study Procedure}
    \label{fig: study design}
\end{figure}
\section{Formative Study: Method}\label{study1}
As discussed in Section~\ref{relatedwork}, there is a lack of understanding regarding whether and how AI-based technologies should be designed to meet caregivers' information needs while their older adult patients are in the ICU. To address this gap and identify key design requirements, we conducted a two-stage study, as illustrated in Fig~\ref{fig: study design}.
At the first stage, we carried out a formative study with 15 caregivers using semi-structured interviews~\cite{longhurst2003semi}.
The study aimed to gather insights on (1) the challenges caregivers encountered in accessing and understanding clinical information in the ICU and (2) their user experience with existing technologies.


\subsection{Participant Recruitment}
To recruit and screen qualified participants, we used convenience sampling~\cite{etikan2016comparison} and designed a survey to collect demographic information and caregiving background details. The survey was distributed via social media platforms, including Twitter, Instagram, and Facebook. Exclusion criteria included individuals who were under 18, not U.S. residents, non-English speakers, or not close family members of older adult patients.
In the end, we recruited 15 participants with experience as family caregivers involved in the hospitalization of older adult ICU patients. 
Details regarding participant demographics and caregiving backgrounds are presented in Table~\ref{tab:participant_demographics}.
This study received approval from the Institutional Review Board (IRB) at the first author’s institution.

\subsection{Interview Procedure}

The interviews were conducted remotely via Zoom and lasted approximately 40–60 minutes.
At the beginning of the interview, we inquired about participants' verbal consent for audio recording and transcription.
During the interviews, we asked participants to recall their experiences as family caregivers when their older family member was admitted to the ICU. We also asked them not to share any identifiable personal information.
The interviews focused on the clinical information caregivers sought to access and understand, as well as the challenges they faced.
We also asked about any technologies they used for accessing and understanding clinical information.
Building on the experiences participants shared, we also asked participants to discuss their usage of any technologies that helped them access and understand clinical information. 
At the end of the interview, we thanked the participants for their time, and each participant received a \$20 Amazon digital gift card as compensation.
The detailed semi-structured interview protocol can be found in Appendix~\ref{Study1script}.
 
After each interview, the audio was transcribed.
Two researchers then independently coded the transcripts using an open coding approach~\cite{corbin2014basics}. They then discussed and reconciled their initial coding schemas, iteratively refining them for clarity and consistency.
The finalized codebook (Appendix ~\ref{Appendix:codebookofformative}) was applied across all transcripts by both researchers, with any discrepancies resolved through team discussions until a consensus was reached.

\begin{table}[h]
\caption{Demographics of Participants and Their ICU Experiences}
\label{tab:participant_demographics}
\begin{tabular}{c|c|c|c|c}
\hline
\textbf{P\#} & \textbf{Gender} & \textbf{Age} & \textbf{Relationship to Patient} & \textbf{Patient Age} \\ \hline
P1  & M  & 26–35 & Father        & 65+  \\ \hline
P2  & M  & 26–35  & Mother         & 65+  \\ \hline
P3  & M  & 26–35  & Grandmother    & 65+  \\ \hline
P4  & F  & 18–25 & Relative       & 80+  \\ \hline
P5  & F  & 26–35 & Mother        & 65+  \\ \hline
P6  & M  & 26–35  & Father         & 67   \\ \hline
P7  & F  & 18–25 & Step-Grandmother & 73  \\ \hline
P8  & M  & 26–35 & Father         & 65   \\ \hline
P9  & M  & 36–45  & Uncle          & 75   \\ \hline
P10 & M  & 18–25  & Father         & 65   \\ \hline
P11 & F  & 26–35 & Uncle          & 65   \\ \hline
P12 & M  & 26–35  & Grandmother    & 65   \\ \hline
P13 & M  & 18-25  & Grandmother    & 70   \\ \hline
P14 & M  & 26–35  & Grandmother    & 68   \\ \hline
P15 & M  & 18-25 & Grandfather    & 72   \\ \hline
\end{tabular}
\end{table}

\section{Formative Study: Results}
In the following section, we detail the challenges caregivers face in accessing and understanding clinical information within the ICU setting.

\subsection{Caregivers' Challenges in Accessing Clinical Information}

\subsubsection{Staying informed about patients' changing status}\label{c1}
While older adult patients are admitted to the ICU, caregivers prioritize staying informed about their daily status. Some even juggle work responsibilities while making trips to the hospital for updates, as P4 noted~\textit{`` it's hard to get to know who's on the shift that day, and sometimes you have to go there physically [to know his situation] ''}.
Beyond the current status, caregivers are also concerned with tracking changes in the patient’s condition from the previous day. Since older adult ICU patients' health can fluctuate drastically, family caregivers seek consistent clinical updates. 

Many interviewees expressed frustration about their limited access to patients' information, which relied solely on verbal updates from the ICU healthcare team. 
This lack of direct access often left caregivers feeling~\textbf{\textit{``left in the dark''}}:
\begin{quote}
\textit{``There were times when I had to seek out doctors or nurses for updates,... I sometimes felt that my patients were seen as an interruption to the busy ICU schedule... In cases where I didn't get information, I didn't get to know what was going on,... I just felt I was ~\textbf{left in the dark}''}(P1).
\end{quote}



\subsubsection{Struggling to align treatment plans and goals with healthcare team}\label{c2}
An extreme yet common issue reported by caregivers is their lack of access to clinical information about the treatments the patients were receiving in the ICU. 
This gap in information access leaves many caregivers feeling excluded from critical decisions about their loved one’s care, creating a sense of helplessness and frustration. 
While most participants expressed trust in the clinicians' expertise and judgment, some had experienced situations where they were only informed after treatments had been administered, leaving no room for questions or understanding. 
P11 articulated this frustration: ~\textit{``I was kind of curious to know the name of the drugs that [were] taken at the moment, but then I think I wasn't given the opportunity to know''}. This lack of access to treatment procedures and plans, even for minor details, can undermine caregivers' confidence in patients' recovery and intensify their anxiety. In life-threatening situations, the lack of clear information on ongoing treatment plans can heighten caregivers' concerns, especially when their loved ones undergo critical treatments (e.g., being connected to life-sustaining machines).

\begin{quote}
    \textit{``[Clinicians did] all the tests, the procedure - [but] what they tested and what they saw, what they got, that kind of deep parts - [I don't know]. I wanted to know the whole [situation] about it, but I think they didn't open it all up, and they were just telling us the diagnosis.''}(P7).
\end{quote}

Another challenge in information access is the treatment goals set by the healthcare team. Caregivers sometimes receive only a general overview of treatment options but lack insight into major care interventions. As a result, they have to piece together scattered information to access the overall treatment goal.
\begin{quote}
~\textit{``so they told me that,... We should probably be more than a week at the ICU and just informing me about... like they might have to maybe perform another surgery if he... doesn't get responsive, and just telling me other treatment options, like ultrafiltration that the end add transplants that might be considered if it isn't responsive.''}(P6)
\end{quote}



\subsection{Caregivers' Challenges in Understanding Clinical Information}
\subsubsection{Interpreting medical terms in clinical report and conversations}\label{c3}

The concept of being ``lost in translation'' is often discussed in relation to the medical term caregivers encounter. From our interviews, we identified two primary sources contributing to this challenge, which frequently leads to information overload: verbal communication with clinicians and the interpretation of clinical reports.

While conversations with healthcare providers often involve complex and unfamiliar medical term, caregivers typically try to seek clarification from doctors. As P4 remarked,~\textit{``some doctors tend to use ambiguous terms''}. In many cases, caregivers feel compelled to ask clinicians for further clarification to make sense of the information shared with them. P5 described her experience:~\textit{``We couldn't understand some medical terms. So we asked him (the doctor) to use a clear word where he explained everything that is happening around...''}.
However, some participants did not have the opportunity to seek clarification from clinicians, making it even more challenging to grasp critical information.

A comparable yet even greater challenge lies in medical reports, where technical content often leaves caregivers confused, prompting them to seek external resources for clarification. Lab test results, in particular, pose a significant challenge, as they are one of the most frequently provided pieces of medical information to caregivers. However, their complex nature makes them difficult to interpret without proper context or guidance.
\begin{quote}
\textit{``when it comes to having to pick reports,...the report is not actually something I could compliment, because there are times there that I do have to... use Google to actually go through and know what it actually means.''}(P8) 
\end{quote}

Caregivers often turn to search engines for quick understanding of medical information, such as definitions of terms or procedural details.
However, without contextual knowledge of the patient’s status, the relevance and quality of the information received varied significantly, often leaving caregivers frustrated and confused.

\begin{quote}
    \textit{ ``When I search online, sometimes the information shows] conflict... It really [just] shows minor information... the equipment used, visiting hours, the policies, it's just like minor information.''}(P2)
\end{quote}


\subsubsection{Asking meaningful questions with healthcare team}\label{c4}

Caregivers also struggle to engage in meaningful conversations with the ICU healthcare team when asking questions. These interactions can often feel unproductive, as caregivers may repeatedly ask broad or unclear questions and receive vague or insufficient responses from healthcare providers in return.
For instance, P1 shared the experience as feeling like ``running in circles'':~\textit{"At some point,...When I asked, [they just said] it is okay, we are doing our work, and his vital signs are okay, and they just tell me I should relax, not to bother myself about it..."}. 
In some cases, caregivers even refrain from asking questions due to negative past experiences, and such interactions discourage open conversation and can prevent caregivers from understanding clinical information about the patient’s care.


To navigate these challenges, some caregivers adopt proactive strategies, such as preparing their questions in advance. This approach helps them feel more confident and ensures they make the most of their limited opportunities to have direct conversations with healthcare providers. For example, as P7 illustrated: \textit{``[I'll prepare] my question, depending on what I see the day I visit her... I'll write what I would like to ask the doctor. So if I go the next day, I can just walk straight to the doctor's office and ask the question."}

\subsection{Design Guidelines: Translating Caregiver Needs into Action}~\label{design space}
Based on the challenges shared by caregivers, several key areas emerge where technology can be improved to better support their needs in accessing and understanding clinical information. From these insights, we derived four design guidelines to shape our prototype: structuring daily patient updates with a timeline, providing accessible clinical information across three key aspects of care, enhancing medical term understanding with context-aware support, and supporting caregivers in structured and informed clinical discussions.

\begin{itemize}
\item \textbf{DG1: Structuring daily patient updates with a timeline:} 
As our formative study results suggest, caregivers need more than just daily patient updates—this aligns with findings from previous research~\cite{mcgonigal2020providing}. They also require comparisons to the previous day (Section~\ref{c1}) to reduce reliance on fragmented verbal updates for accessing clinical information. To support this need, a structured timeline of the patient’s daily updates is essential for helping caregivers better understand the patient’s status throughout their ICU stay.

\item \textbf{DG2: Providing accessible clinical information across three key aspects of care:}
Our interview study also highlighted caregivers' focus on three key aspects of clinical information: treatment plans, treatment goals (Section~\ref{c2}), and lab test results (Section~\ref{c3}). Additionally, this information should be presented in a unified and easy-to-understand manner.

\item \textbf{DG3: Enhancing medical term understanding with context-aware support:} 
Moreover, medical term encountered in different situations can confuse caregivers, whether in conversations with healthcare providers or in test reports, regardless of their ability to seek clarification from the professional healthcare team (Section~\ref{c3}). Additionally, searching for information online without the context of the patient’s condition can lead to even greater confusion. Thus, a context-aware chatbot that serves as an always-available resource can facilitate caregivers' personalized information inquiries and improve their understanding.

\item \textbf{DG4: Supporting caregivers in structured and informed clinical discussions:}
Caregivers' proactive engagement emerged as a key theme in our findings. Despite limited opportunities, they actively seek conversations with the healthcare team (Section~\ref{c1}). One effective practice is preparing a meaningful list of questions in advance. Thus, a feature designed to suggest context-relevant questions can help caregivers better understand the patient’s condition and provide a structured framework for discussions (Section~\ref{c4}).
\end{itemize}

\section{\projectname: an AI-Based Prototype for Supporting Caregivers of Older Adult ICU Patients }
\revise{Building on the design space outlined in Section~\ref{design space}, we developed an AI-based prototype system, \projectname, to support caregivers in accessing and understanding clinical information. An overview of the system architecture is shown in Figure~\ref{teaser}. The protoype system is designed to seamlessly collect patient data from the EHR system, process it through the back-end structure (Section~\ref{sec:backend}), and deliver the resulting information through the front-end interfaces to family caregivers (Section~\ref{sec:frontend}).}

\subsection{Front-end Interface Design}\label{sec:frontend}
The front-end interface includes two interconnected components: a visual timeline of patients' medical events and an LLM-based chatbot to facilitate personalized information inquiries through conversations. 
\begin{figure}
    \centering
    \includegraphics[width=1\linewidth]{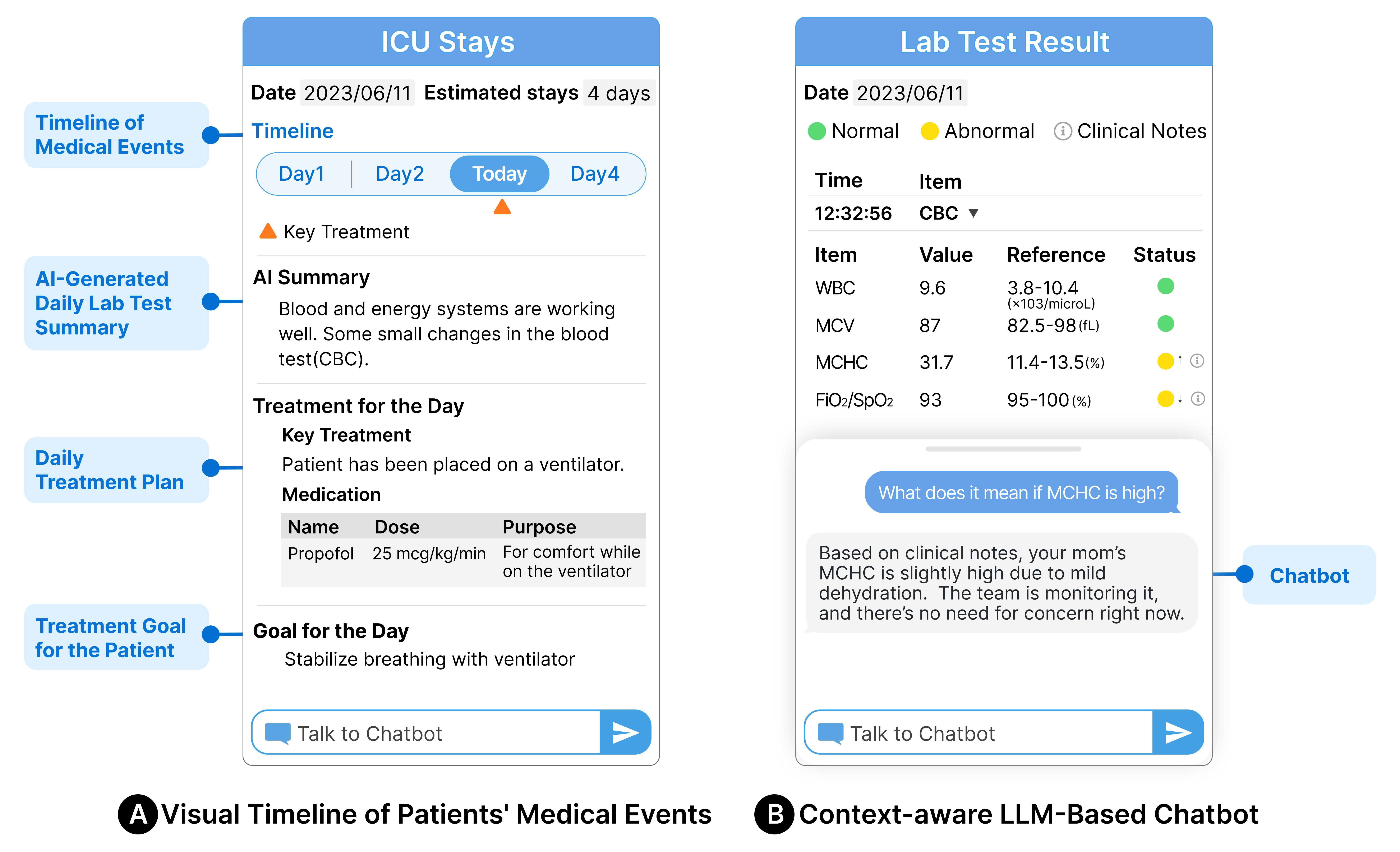}
    \caption{\projectname Design: (A) A visual timeline that maps a patient's medical events, featuring AI-generated summaries of lab test results, key treatment plans, and daily goals to provide a clear overview of the patient’s status. (B) An example of a context-aware LLM-based chatbot that provides insights into medical reports, helping users interpret test results and identify trends.}
    \label{fig:Overview}
\end{figure}

\subsubsection{Visual Timeline of Patients' Medical Events}
To help caregivers better \textbf{\textit{access}} patients' conditions, this interface transforms fragmented clinical data into a structured report, organized in a timeline format (as shown in Fig.~\ref{fig:Overview} (A)). 
The top section features a sequential timeline highlighting key treatment milestones (e.g., ventilator placement). By selecting a specific day, caregivers can track and compare clinical reports to monitor the patient's ICU progress (\textbf{DG1}). Below, a daily overview organizes information into three key modules that caregivers primarily focus on regarding the patient (\textbf{DG2}), including: 

~\textbf{The AI-generated summary} module takes structured lab test results as input and generates plain-language explanations to help caregivers make sense of medical values and trends. The underlying LLM-based summarization model continuously updates the summary when new lab results are available, helping caregivers stay informed about the patient’s physiological changes.

~\textbf{The Treatment for the Day} and ~\textbf{Goals for the Day modules} take unstructured clinical notes as input and extract relevant content using an information extraction module. Specifically, the Treatment module identifies key interventions and medications and generates brief explanations of their purpose. The Goals module highlights actionable priorities in the patient’s care plan, outlining next steps that caregivers can track. Together, these modules help caregivers understand not only what is being done but also why, enabling them to stay aligned with ongoing care decisions.

\subsubsection{Context-Aware LLM-Based Chatbot}

\begin{figure}
    \centering
    \includegraphics[width=1\linewidth]{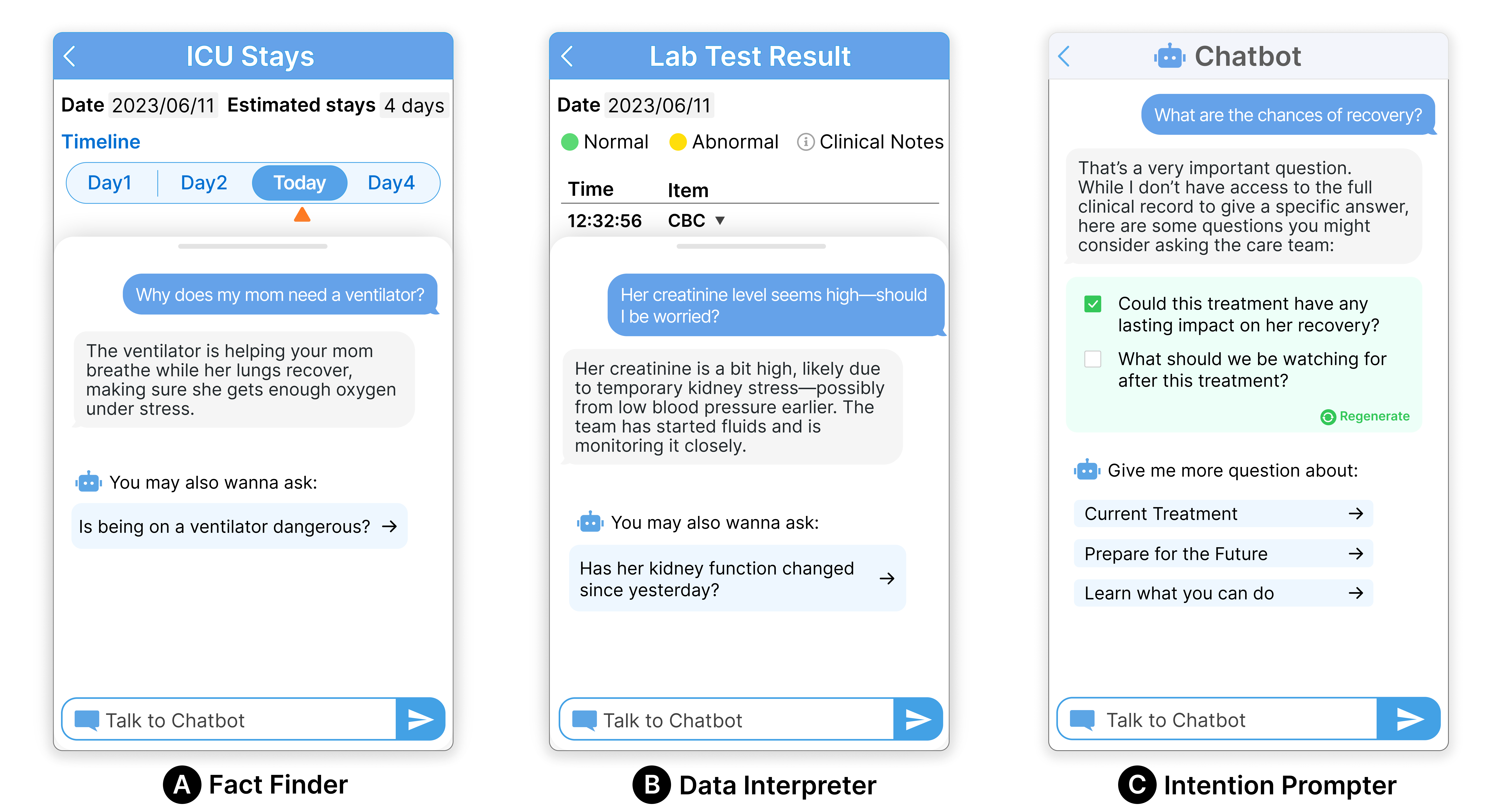}
    \caption{Context-Aware LLM-Based Chatbot Design: (A) Fact Finder – Responds to user queries about ICU treatments and suggests relevant follow-up questions. (B) Data Interpreter – Explains lab test results and clinical data in an accessible manner. (C) Intention Prompter – Assists users in formulating meaningful questions for healthcare providers.}
    \label{fig:chatbot}
\end{figure}

To help caregivers better \textbf{\textit{understand}} clinical information, we designed an LLM-based chatbot that leverages EHR data as part of its information source (as shown in Fig.~\ref{fig:chatbot}). 
Instead of providing generic medical explanations, it tailors responses based on the patient’s specific clinical data. This chatbot enables personalized inquiries through flexible text-based conversations and supports caregivers in three key ways while also suggesting relevant follow-up questions (\textbf{DG3}). The chatbot operates through three core interaction modes:

\textbf{Fact Finder} (Fig.\ref{fig:chatbot} (A)) responds to direct caregiver questions by retrieving relevant clinical context and offering lay explanations—for example, clarifying the need for a ventilator as ``supporting oxygen exchange during lung recovery.'' To encourage proactive advocacy, it also suggests follow-up questions such as ``What are the signs indicating ventilator removal?'', helping caregivers navigate unfamiliar clinical territory.

\textbf{Data Interpreter} (Fig. \ref{fig:chatbot} (B)) takes lab test results and other structured clinical data as input and translates them into caregiver-friendly responses and generates caregiver-friendly explanations that clarify both the meaning of individual values and their implications for the patient’s condition.

\textbf{Intention Prompter} chatbot (Fig.~\ref{fig:chatbot} (C)) is designed to support caregivers when the chatbot lacks access to specific clinical information or when caregiver questions are too vague to generate a precise response. In these situations, the chatbot offers proactive guidance by helping caregivers prepare questions for upcoming discussions with clinicians (\textbf{DG4}). It presents categorized follow-up prompts across three thematic areas: ``Current Treatment,'' which encourages caregivers to ask about the patient's ongoing treatments and medical status, ``Prepare for the Future,'' which helps them anticipate potential care decisions and long-term planning, and ``Learn What You Can Do,'' which guides them in inquiring about actionable steps they can take to provide better support. 

\subsection{Backend Structure Design}\label{sec:backend}


\revise{The back-end architecture of \projectname stores, processes, and analyzes clinical data collected from Electronic Health Records (EHRs) to support the two front-end components described in Section~\ref{sec:frontend}.
It consists of three core modules: the Data Extraction Module, the Summarization Module, and the Conversational Module.}

\revise{Our back-end algorithms are inspired by prior work on patient-centered EHR interaction~\cite{zhang2023ehr} and longitudinal clinical reasoning using LLMs~\cite{kruse2025zero}. For example,~\citet{kruse2025zero} systematically evaluated state-of-the-art LLMs on multi-day Assessment \& Plan generation and discharge summarization using MIMIC-III data, benchmarking three prompting strategies—direct generation, Retrieval-Augmented Generation (RAG)~\cite{gao2023retrieval, lewis2020retrieval}, and Chain-of-Thought (CoT)~\cite{wei2022chain}. Inspired by this approach, our \textbf{Data Extraction Module} leverages LLM-based processing to standardize and structure EHR data (e.g., lab tests, medications, vital signs) and temporally order clinical events to support retrieval and generation, while our \textbf{Summarization Module} uses LLMs to retrieve multi-day evidence and perform temporal reasoning across structured and unstructured data to produce daily summaries.}

\revise{Complementary to this,~\citet{zhang2023ehr} developed the NoteAid EHR Interaction Pipeline, which leverages LLMs — specifically GPT-3.5-Turbo and GPT-4 — to construct a synthetic dialogue dataset for patient education on EHR notes. Their system pairs a Mock-Patient LLM agent and an Assistant LLM agent to simulate multi-round conversations (three rounds each) that perform Q\&A and explanation tasks on unstructured discharge notes from MIMIC-III~\cite{johnson2016mimic} and MADE~\cite{jagannatha2019overview}. This approach yielded 43{,}504 synthetic interaction instances, demonstrating the potential ofLLMs to enhance patients’ understanding of their medical records . Building on this paradigm, our~\textbf{Conversational Module} similarly grounds its responses in patient charts and produces plain-language explanations tailored for caregivers.}

\revise{To further assess the clinical validity of the AI-generated outputs, we consulted two experienced board-certified clinicians to conduct an in-depth evaluation. We selected four pairs of AI-generated summaries and their corresponding EHR inputs from \citet{zhang2023ehr}’s example. The clinicians rated each summary on five criteria—\textit{Overall Accuracy, Hallucination, Readability, Specificity}, and \textit{Overall Quality}. The detailed questionnaire used for this evaluation is provided in Appendix~\ref{questionnaire}.}

\begin{figure}
    \centering
    \includegraphics[width=0.5\linewidth]{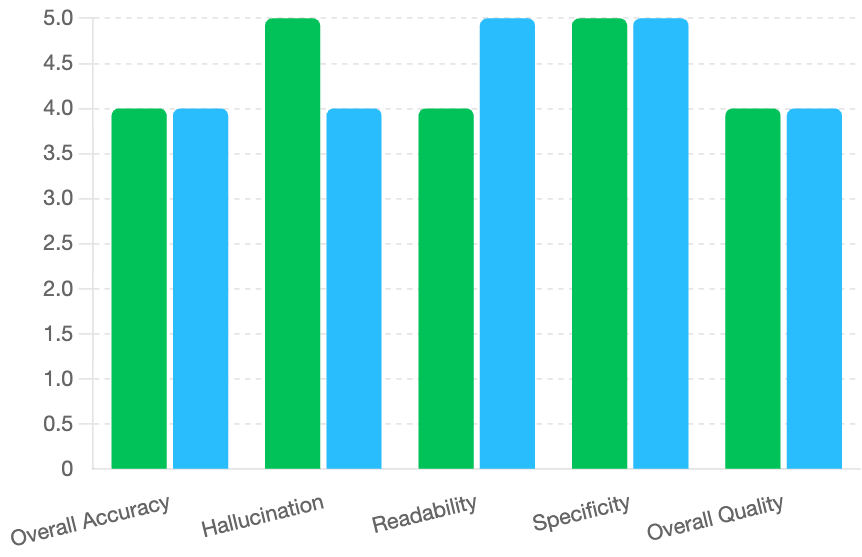}
    \caption{Evaluation scores given by two clinician experts (C1 and C2) across five criteria—Overall Accuracy, Hallucination, Readability, Specificity, and Overall Quality—for the AI-generated outputs.}
    \label{fig:expert_eval}
\end{figure}

\revise{The results of this evaluation are shown in Figure~\ref{fig:expert_eval}. As illustrated, the clinicians gave consistently high ratings for the overall quality of the AI-generated outputs.(\textit{M} = 4), including \textit{Overall Accuracy} (\textit{M} = 4), \textit{Hallucination} (\textit{M} = 4.5), \textit{Readability} (\textit{M} = 4.5), and \textit{Specificity} (\textit{M} = 5).}

\section{Prototype Evaluation}
\subsection{Participants and Recruitment}

\revise{For the second stage of our study, we recruited the same ten caregivers (P6-P15) from Section~\ref{study1} to conduct a usability evaluation of our prototype.
We also recruited 2 clinicians with significant experience in this area by convenience sampling~\cite{etikan2016comparison}, as detailed in table~\ref{tab:clinicians}. Recruitment involved leveraging professional networks and referrals from colleagues.
Each prototype evaluation session was conducted remotely via Zoom and lasted approximately 30–45 minutes.
To comply with usage regulations requiring approval for accessing MIMIC-III~\cite{johnson2016mimic} data, we integrated de-identified, synthesized patient data into our prototype. In this paper, we present only mock data to illustrate the user interface.}

\begin{table}[t]
\centering
\small
\begin{tabular}{lllll}
\toprule
\textbf{C\#} & \textbf{Gender} & \textbf{Department} & \textbf{Job Title} & \textbf{Year of Practice} \\
\midrule
C1 & Female & Critical Care & Physician & 13 Years  \\
C2 & Female & Emergency Medicine & Physician & 10 Years\\
\bottomrule
\end{tabular}
\caption{Clinician experts’ background summary.}
\label{tab:clinicians}
\end{table}

\revise{All sessions were recorded and transcribed with participant consent. For each design prototype, we allocated approximately 10–15 minutes for family caregivers to explore its features. After a brief introduction to the interface, caregivers were encouraged to interact with the system, share their impressions, and evaluate its perceived usefulness in accessing and understanding a patient’s status. They also reflected on how they might have used the tool to obtain relevant medical information, identified the most helpful aspects, and suggested potential improvements to enhance overall usability. In parallel, two clinical experts reviewed the same prototypes. They were asked to examine the AI-generated outputs (daily summaries, treatment plans, and chatbot responses) for \textit{Overall Accuracy, Hallucination, Readability, Specificity}, and \textit{Overall Quality} (as detailed in Section~\ref{sec:backend}), and to provide usability feedback on the interface from a clinical perspective.}

\revise{The detailed semi-structured interview protocols are provided in Appendix~\ref{Study2script}. Following the interviews, two researchers independently analyzed the transcripts using an inductive approach~\cite{thomas2006general}, identifying key insights related to usability and potential design improvements for each prototype; the final codebook is provided in Appendix~\ref{Appendix:codebookofforevaluation}.}

\subsection{Result}
\revise{In the following section, we present the results of the usability evaluation, which captures valuable feedback from family caregivers and clinician experts on our prototype.
Overall, family caregivers greatly appreciated the prototype’s ability to enhance their access to clinical information by providing consistent, timely, and easy-to-understand updates.
Meanwhile, they found the prototype’s capability to integrate with personalized patient EHRs particularly helpful for them to understand clinical information, as it could effectively respond to their specific inquiries, uncover additional relevant details, and offer deeper insights into the patient’s condition.
In parallel, clinicians provided positive feedback on its potential to support accurate information delivery and to help caregivers interpret complex medical data in a patient-specific context.}

\subsubsection{Visual Timeline of ICU Stay: Supporting Caregivers’ Access to Consistent Information}

\paragraph{Synthesizing Fragmented Information into a Cohesive Timeline}
Overall, caregivers found the visual timeline view to be a valuable tool for better supporting their access to dynamic patient status changes during ICU stays.
Building on findings from the formative study (See Section~\ref{c1}), which highlighted caregivers' difficulties with inconsistent information structuring and accessibility in ICU workflows, the timeline interface is designed to address this challenge. Caregivers often rely on fragmented communication with healthcare providers, making it difficult to piece together a comprehensive understanding of the patient’s condition and treatment process.
\revise{A unified interface that uses a timeline to visualize patient progress in the ICU over time could address this gap by enabling them to track trends and identify critical events of the patient's ICU stay. also provided comments about its potential to help families recognize the passage of time during the ICU stay. }

\begin{quote}
\revise{\textit{`` (this) might provide some relevant time to the family,... because I think sometimes the families don't realize how long it has been.''}(C1)}
\end{quote}
The timeline was useful not only for individual caregivers but also as a shared communication tool for the patient’s family.
Some caregivers noted that it helped bridge information gaps between family caregivers and other family members, making it easier to share updates on the patient’s condition.

\begin{quote}
~\textit{``It's like every time I want to check out the progress or every time I want to send some results to my siblings, I would try to use the tab [on the switch of different days] and just saying, this is this progress chart [shows the patient's condition.''}(P6)
\end{quote}

\revise{Additionally, the timeline helped reduce information overload — a common challenge in ICU environments — by organizing complex clinical data into a clear, time-based summary. As patients often undergo many procedures each day, as commented by clinician:}

\begin{quote}
    \revise{\textit{``My patients have many procedures, and it's really nice to have a tab that says: on the first day after the operating room, they had a central line placed. (On) the second day, they had a tracheostomy. Those things are actually quite nice to [be] organized.''}(C2)}
\end{quote}

This design minimized caregivers' cognitive effort, allowing family caregivers to focus on understanding the patient's overall status rather than manually piecing together fragmented updates. The interface supported on-demand access to critical information, enabling caregivers to check in as needed rather than constantly monitoring data:

\begin{quote}
\textit{``[I] might not be daily [to check on that information], but anytime you want to really know about what is really wrong, to have more understanding [on the patient's status], you just look at it.''}(P7)
\end{quote}

\paragraph{Balancing Simplicity and Depth in AI-Generated Summaries} 
AI-generated summaries of lab test results effectively provided family caregivers with a quick overview of the patient’s status.
However, both clinicians and caregivers emphasized the need to balance simplicity with sufficient medical context. While summaries were found valuable for gaining an initial understanding—especially in high-stress ICU environments where cognitive load is a concern—they also expressed concerns about oversimplification. 
Reducing complex medical information to brief summaries risks omitting critical details, making it harder for caregivers to fully grasp the care process and provide effective support.
Since older adult patients often have multiple chronic conditions before being admitted to the ICU, caregivers prefer more than just AI-generated summaries of daily lab test results. Incorporating a patient’s prior medical history into these summaries is crucial, as it would help caregivers better understand acute conditions and how they relate to the patient’s overall health.

\begin{quote}
\textit{``Any [other] information I would like to see is the previous illness of the patient... if the patient is having any other illness [that leads to this result].''}(P10)
\end{quote}

\paragraph{Involving Caregivers in Treatment Plans and Goals}

\revise{For caregivers who cannot always be physically present in the ICU, access to detailed information about their loved one’s daily treatment plans and goals is crucial for maintaining a sense of connection and involvement in care.
While our prototype outlines treatment plans—including key milestones, medications, dosages, and their purposes—to help caregivers stay engaged, many participants emphasized the need for even greater transparency to further enhance their involvement in the care process. Moreover, as mentioned in Section~\ref{sec:backend}, clinicians expressed slight concerns about readability (M = 4.5). Specifically, they emphasized the need to clarify which treatment plans should be prioritized when presented to caregivers: }

\begin{quote}
    \revise{\textit{... [the treatment plan] gonna be really challenging to be put into a summary for family caregivers, most of these patients are on ten plus medications..}(C2) }
\end{quote}


Beyond clarity, caregivers also seek insight into the broader arc of care, including how treatments correlate with patient progress. P9 highlighted this need, noting that understanding expected outcomes helps contextualize daily interventions:~\textit{``Yeah. It[the goals] like the kind of results, [and improvements] that we should be expecting, that's the only thing I noticed''}. Similarly, P10 stressed the value of linking treatments to patient responses:~\textit{``[besides providing the treatment, I want to know] how he is responding to treatment, [based on] the kind of treatment has been given that day, and the activities of the day''}.

\subsubsection{Context-Aware LLM-Based Chatbot: Enhancing Caregivers’ Understanding of ICU Information}




\paragraph{Providing Reliable, On-Demand Responses}
Caregivers prioritized the need for accurate, instant explanations of ICU equipment and treatments—a demand met by the chatbot’s Fact Finder feature. 

By providing immediate explanations of medical treatment plans and goals, the chatbot helps caregivers understand care decisions without solely depending on clinician availability. For example, when asked about ventilator use, the chatbot explains how the device stabilizes oxygen saturation and supports lung healing, making complex medical concepts more accessible and reducing delays in understanding essential information.

\begin{quote}
\textit{`` It helps because there are times ... instead of waiting to see a doctor to actually explain [the treatment plan], you could actually just use the [chatbot] in this portal to know the reason why a particular equipment or a particular thing is actually administered.''}(P8)
\end{quote}

While caregivers generally appreciated the chatbot, some raised concerns about verifying its responses and ensuring accuracy. For instance, P15 highlighted the importance of cross-checking AI-generated information with medical professionals, stating:~\textit{``I wouldn't like one hundred percent trust the chatbot that answers my question. But I think with time, I might verify [the] answers [by] speaking to a doctor.''}

\paragraph{Translating Complex Data into Clear Insights}
\revise{Medical data in professional reports, often dense with jargon and numerical thresholds, can be overwhelming for caregivers unfamiliar with clinical term. Caregivers in our study expressed a need for clearer explanations of medical references to better understand their significance. The Data Interpreter helps address this challenge by providing definitions and contextual insights.
Clinicians appreciated its ability to help caregivers address complex questions about lab test results, while expressing slight concerns about potential hallucinations in numerical interpretation (M = 4.5), as mentioned in Section~\ref{sec:backend}: }
\begin{quote}
    \revise{\textit{``.. because we'll be in the middle of rounds, and families will ask us very specific questions (like) I can see my family members' results constantly, and they're asking about every little number, including numbers I don't even look at, (So I tell them) it doesn't matter...''}(C1)}
\end{quote}


Caregivers also valued the chatbot’s ability to cross-check and validate information by referencing the patient’s medical history, rather than merely summarizing or explaining test results in isolation. This feature helps them interpret medical data more accurately, reducing unnecessary anxiety over abnormal test results by providing context and a clearer understanding of their significance.

\begin{quote}
    \textit{``I think having past diagnosis [connected to this chatbot] would be helpful... [then] you know [the] diagnose [is] proper, [and] without having any side effects [to the patients]''}(P15)
\end{quote}


\paragraph{Structuring Conversations and Improving Health Literacy}
\revise{Beyond providing reactive support, the Intention Prompter helps caregivers anticipate and navigate conversations with the clinical team more effectively. Clinicians also emphasized that these questions should be tailored based on the patient’s specific condition:~\textit{``...if they brought it to me with the context of their own ideas and interpretation of what's been going on, it could be helpful...''}(C2).}
By suggesting questions based on the patient's status, the system helps move conversations beyond repetitive exchanges, making them more meaningful and informative. 

\begin{quote}
\textit{``Instead of racking my brain on what to ask the doctor,... I'll just like go straight to the point instead of trying to find things out, I'll just ask doctors the possible questions that I got from [the chatbot].''}(P12)
\end{quote}


This feature also helps caregivers improve their health literacy by guiding them to identify key topics for discussion with clinicians in advance. By preparing them with relevant information beforehand, it ensures they retain critical details during consultations and feel more confident in their conversations.
P6 emphasized its role in reducing cognitive overload, stating:~\textit{``Just the questions [listed] here would help me prepare questions [for the conversations], because the doctor mentioned so many things that I wouldn't even recall [and] I wouldn't even understand. But if I had prepared my question before and after [that] the doctor answered, then it helped me understand the situation''}. 

\section{Discussion}
Drawing from these findings, we begin by exploring how our design can reposition and empower family caregivers as active participants in ICU care, rather than passive or ancillary stakeholders  (Section \ref{D1}).
Next, we outline key design considerations for future systems that support family caregivers' information needs in clinical settings for older adult patients(Section \ref{D2}).
Finally, we highlight risks and ethical concerns associated with the use of technology and AI in real-world implementations (Section \ref{D3}).
Finally, we present the limitations of our work and suggest directions for future research (Section \ref{D4}).

\subsection{Beyond Ancillary Roles: Empowering Caregivers as Informal Healthcare Partners}\label{D1}
Traditionally, in clinical settings, efforts to improve the quality of healthcare provision have centered around patient-centered care, which emphasizes increasing patient engagement in accessing clinical information and participating in communication throughout the care process~\cite{baker2001crossing, davis20052020}.
Within this framework, family caregivers have often been characterized as proxies or supplementary stakeholders, positioned primarily as assistants to clinicians or patients rather than active contributors to care~\cite{Foong2020, Bhat2023, essen2004proxy}.
Our formative study (see Section~\ref{study1}) further highlights how family caregivers often feel excluded—or even~\textbf{\textit{``left in the dark''}}—in ICU settings when trying to access and understand clinical information. 
In the ICU, caregivers shift from peripheral roles to active participants, who need to advocate for patients and make critical decisions on their behalf~\cite{au2017family}.
Recognizing the critical role of caregivers' involvement, recent healthcare research has increasingly acknowledged the importance of integrating them into clinical settings, proposing frameworks such as family-centered care (FCC)~\cite{jolley2009evolution, care2012patient, dunst1996empowerment} and practices to institutionalize their participation. 

In parallel, research in Computer-Supported Cooperative Work (CSCW) and Human-Computer Interaction (HCI) has adopted an ecological perspective to investigate the socio-technical ecosystems surrounding family caregivers in clinical settings. 
Empirical studies have systematically mapped caregivers’ multifaceted roles in clinical settings - from providing emotional support to coordinating appointments, transportation, and daily care during hospital stays~\cite{Miller2016, Pina2017, Barbarin2015}. 
These findings collectively emphasize the importance of designing systems and practices that can better support caregivers’ evolving roles within complex clinical environments.
\revise{Our research seeks to complement this body of work by further unpacking the complexities of caregivers’ informational needs as primary stakeholders, particularly within the high-stakes environment of the Intensive Care Unit (ICU).
Moreover, with the add-on feedback from clinicians, they recognized the challenges faced by caregivers and how our system can help bridge informational gaps and support their understanding of patients’ conditions.
Specifically, we highlight the importance of two-way information flow and adaptive support systems that accommodate caregivers' varying levels of health literacy.
Our findings show that technology can empower caregivers to take on a more active role, not just as recipients of information, but as engaged participants in care coordination. }

While our work demonstrates the potential of technology to mitigate caregiver exclusion, it also surfaces critical tensions in scaling these interventions. 
For example, while AI-based continuous patient-status tracking is essential in critical care, it may unintentionally increase caregiver mental strain in chronic care settings, where long-term stability is the primary concern rather than constant changes~\cite{Bhat2023}.
These technologies should not be applied uniformly across various clinical settings, and we argue against one-size-fits-all solutions. 
Instead, technologies should adapt to the context of caregiving: urgent situations require real-time updates, while long-term care may benefit more from tracking clinical trends over time.
In a word, technology should be designed with input from caregivers and a focus on their values, especially in situations where collaboration with clinicians is crucial, such as dementia care \cite{smriti2024emotion} and post-operative recovery \cite{Kaziunas2015}.

\subsection{Design Implications for Future Systems}\label{D2}


Based on insights from the formative study, we established a set of design goals to guide the development of an AI-based, caregiver-centered clinical information prototype (See Section~\ref{design space}). 
In this section, we propose design implications for future systems that support caregivers in clinical settings, drawing on insights from Human-AI Collaboration and Computer-Supported Cooperative Work (CSCW). 

\subsubsection{From the perspective of Human-AI Collaboration} Accessing and understanding clinical information is one of the most cognitively demanding tasks for family caregivers, especially in the high-pressure, fast-paced environment of the ICU. 
\revise{Based on findings from our two-stage study, we explored with caregivers how AI could help streamline their processes of accessing and making sense of information. A key direction for future design is to go beyond basic summarization or extraction. }
Instead, AI systems should engage in adaptive, ongoing interactions that learn from caregivers’ responses and progressively personalize information delivery.

In addition, AI systems deployed in clinical settings should be designed to foster appropriate user trust and expectations. Given the complexity of these systems and the high-stakes nature of ICU environments, it is critical to clearly communicate the system’s capabilities and limitations to caregivers. 
\revise{While clinicians generally gave high overall ratings for the AI-generated outputs, caregivers in our study expressed concerns about the ambiguity of these summaries, particularly when numerical data were presented without adequate explanation.}
We suggest that designers offer explanations tailored to the underlying technologies used in the system, such as clarifying how outputs were generated and presenting insights in concise, easy-to-understand formats.

\subsubsection{From the perspective of CSCW}
ICU caregiving is both collaborative and isolating—while families play a crucial role, they often feel excluded from clinical workflows structured around the ICU’s fast-paced environment.  
CSCW research explores ways to improve teamwork in healthcare by developing systems that help caregivers and healthcare providers share and interpret important information across different roles and data sources~\cite{ziqi2024, Yuexing2024, Bowers2024}. 
In our study, caregivers expressed a desire to supplement clinical data with their firsthand insights and their need for greater transparency and inclusion in coordinated care. 
To address this, future systems could allow caregivers to contribute their observations (e.g., concerns about specific treatments), with both clinicians and AI validating the information. 
This approach would facilitate structured information exchange and strengthen collaboration with healthcare providers.
If caregivers have the opportunity to provide input to the system and receive feedback from AI or healthcare providers, they can develop a better understanding of medical information as well as improve their health literacy. 
\revise{Furthermore, as clinicians intend to use this interface to update patient information, it might bridge the gap between clinical decision-making and caregivers’ understanding, fostering more aligned expectations and collaborative care.}


\subsection{Risks, Technical and Ethical Concerns of AI-based Clinical Information Technologies}\label{D3}

While AI-based technologies hold promise in supporting caregivers' access and understanding of clinical information, certain emerging risks and ethical concerns should also be carefully considered.

\subsubsection{Infrastructure and Technical Challenges}\label{Realtime}
\revise{Successful real-time implementation of \projectname in clinical environments requires coordinated efforts across both infrastructure and healthcare teams. From the perspective of infrastructure, many mHealth applications—such as MyChart~\cite{ramsey2018increasing}—have already been integrated into standard clinical workflows and tightly coupled with mainstream electronic health record (EHR) systems. This trend demonstrates that healthcare institutions possess mature data interfaces, robust security protocols, and established mechanisms for workflow integration. As a result, incorporating the conversational capabilities of \projectname into existing EHRs or remote monitoring platforms is both technically and institutionally viable.}

\revise{Equally important is the sustained engagement of healthcare teams to ensure the quality and reliability of information, as well as the overall stability of system operations. AI-based clinical technologies can support the interpretation of medical information and the generation of initial responses, but in the high-risk, continuously monitored environment of the ICU, human oversight remains essential to preserve medical accuracy and contextual relevance. Clinical involvement also enhances information credibility and allows for critical judgment in abnormal or emergency situations. In parallel, system deployment must address ethical concerns and workflow integration challenges, including the need to prevent additional cognitive or operational burdens for healthcare team~\cite{wu2024clinical, zhang2024}.}

\subsubsection{Ethical Concerns of AI-based Clinical Information Technologies}
Another central tension emerges from our prototype’s dual role as both a support tool and a potential stressor. 
Caregivers acknowledged that our prototype could facilitate their access to information, such as tracking changes in a patient’s condition, yet they also expressed concerns about information overload, particularly when frequent, non-essential updates increased their cognitive burden. 
This aligns with previous research on AI-supported decision tools, which found that frequent notifications and complex data presentations can inadvertently increase user stress~\cite{wang2023}. 

Beyond cognitive burden, trust in AI-generated insights emerged as a major concern among caregivers. Participants found AI-generated explanations helpful in understanding lab results, but still felt the need to cross-validate responses with healthcare providers before making care decisions. 
This reflects a broader challenge in AI adoption—while AI can enhance information accessibility, caregivers do not perceive it as an authoritative source but rather as a supplementary tool. Prior research on clinical decision support systems has noted similar patterns, where users rely on AI-generated insights but defer final decision-making to human experts~\cite{zhang2024}. This underscores the need for explainable AI (XAI) approaches that offer not only summaries but also data sources, helping caregivers understand why certain answers are made and when they should seek human validation.

\subsection{Limitations and Future Work}\label{D4}

\revise{Our study has several limitations that should be acknowledged. 
First, our participant pool was relatively limited, consisting of 15 family caregivers in the formative study and 10 family caregivers and 2 clinicians in the evaluation phase. While these participants provided valuable insights, the small sample size may limit the generalizability of our findings. Future research should aim to include a larger and more diverse group of caregivers and clinicians to ensure broader applicability.
Nevertheless, our analysis reached a point of saturation, and participants contributed rich and diverse perspectives that were sufficient for the exploratory and proof-of-concept stage of this research. This ensures our study results comprehensively reflect the experiences and perspectives of caregivers for older adult patients in critical care settings.}

\revise{Second, the prototype developed in this study is a proof-of-concept and has not been integrated with Electronic Health Record (EHR) systems or deployed in real-world clinical settings. While the prototype demonstrates the potential of AI-based tools to support caregivers as discussed in Sec~\ref{Realtime}, its effectiveness and usability in practical, high-stakes environments remain untested. Future work should focus on integrating such tools with existing healthcare infrastructures and evaluating their performance in real-world scenarios.}


\revise{Lastly, while our research primarily addresses the ICU setting, there is a broader need to explore how similar tools can be adapted for other clinical environments. The challenges faced by caregivers in ICUs may differ from those in other settings, such as long-term care facilities or home-based care. Future work should delve deeper into these contexts to develop more versatile solutions that can support caregivers across a wide range of clinical scenarios.}

\section{Conclusion}

In this study, we investigated the information needs of family caregivers of ICU older adult patients and designed an AI-based prototype, \projectname,  to address their challenges in accessing and understanding clinical information.
\revise{Through a formative study with 15 caregivers, we first uncovered the multifaceted challenges caregivers face in the ICU. 
Building on these insights, we designed an AI-based prototype and evaluated it with 10 caregivers and 2 clinicians.}
Results from the evaluation study suggested that, despite participants identifying potential trust risks in using an AI-based system, they emphasized that the plain-language explanations and unified visual timeline empowered them to more effectively access and understand information in critical care settings.
This paper contributes to HCI research in healthcare by highlighting the need for AI systems that actively engage caregivers and ensure they feel informed and supported throughout the care process, rather than sidelined.

\bibliographystyle{ACM-Reference-Format}
\bibliography{sample-base}


\begin{thebibliography}{97}


\ifx \showCODEN    \undefined \def \showCODEN     #1{\unskip}     \fi
\ifx \showDOI      \undefined \def \showDOI       #1{#1}\fi
\ifx \showISBNx    \undefined \def \showISBNx     #1{\unskip}     \fi
\ifx \showISBNxiii \undefined \def \showISBNxiii  #1{\unskip}     \fi
\ifx \showISSN     \undefined \def \showISSN      #1{\unskip}     \fi
\ifx \showLCCN     \undefined \def \showLCCN      #1{\unskip}     \fi
\ifx \shownote     \undefined \def \shownote      #1{#1}          \fi
\ifx \showarticletitle \undefined \def \showarticletitle #1{#1}   \fi
\ifx \showURL      \undefined \def \showURL       {\relax}        \fi
\providecommand\bibfield[2]{#2}
\providecommand\bibinfo[2]{#2}
\providecommand\natexlab[1]{#1}
\providecommand\showeprint[2][]{arXiv:#2}

\bibitem[Afshar et~al\mbox{.}(2024)]%
        {afshar2024prompt}
\bibfield{author}{\bibinfo{person}{Majid Afshar}, \bibinfo{person}{Yanjun Gao},
  \bibinfo{person}{Graham Wills}, \bibinfo{person}{Jason Wang},
  \bibinfo{person}{Matthew~M Churpek}, \bibinfo{person}{Christa~J
  Westenberger}, \bibinfo{person}{David~T Kunstman}, \bibinfo{person}{Joel~E
  Gordon}, \bibinfo{person}{Cherodeep Goswami}, \bibinfo{person}{Frank~J Liao},
  {et~al\mbox{.}}} \bibinfo{year}{2024}\natexlab{}.
\newblock \showarticletitle{Prompt engineering with a large language model to
  assist providers in responding to patient inquiries: a real-time
  implementation in the electronic health record}.
\newblock \bibinfo{journal}{\emph{JAMIA open}} \bibinfo{volume}{7},
  \bibinfo{number}{3} (\bibinfo{year}{2024}), \bibinfo{pages}{ooae080}.
\newblock


\bibitem[Al-Mutair et~al\mbox{.}(2013)]%
        {al2013family}
\bibfield{author}{\bibinfo{person}{Abbas~Saleh Al-Mutair},
  \bibinfo{person}{Virginia Plummer}, \bibinfo{person}{Anthony O'Brien}, {and}
  \bibinfo{person}{Rosemary Clerehan}.} \bibinfo{year}{2013}\natexlab{}.
\newblock \showarticletitle{Family needs and involvement in the intensive care
  unit: a literature review}.
\newblock \bibinfo{journal}{\emph{Journal of clinical nursing}}
  \bibinfo{volume}{22}, \bibinfo{number}{13-14} (\bibinfo{year}{2013}),
  \bibinfo{pages}{1805--1817}.
\newblock


\bibitem[Alfheim et~al\mbox{.}(2019)]%
        {alfheim2019post}
\bibfield{author}{\bibinfo{person}{Hanne~Birgit Alfheim},
  \bibinfo{person}{Kristin Hofs{\o}}, \bibinfo{person}{Milada~Cvancarova
  Sm{\aa}stuen}, \bibinfo{person}{Kirsti T{\o}ien}, \bibinfo{person}{Leiv~Arne
  Rosseland}, {and} \bibinfo{person}{Tone Rust{\o}en}.}
  \bibinfo{year}{2019}\natexlab{}.
\newblock \showarticletitle{Post-traumatic stress symptoms in family caregivers
  of intensive care unit patients: A longitudinal study}.
\newblock \bibinfo{journal}{\emph{Intensive and Critical Care Nursing}}
  \bibinfo{volume}{50} (\bibinfo{year}{2019}), \bibinfo{pages}{5--10}.
\newblock


\bibitem[Angus et~al\mbox{.}(2006)]%
        {angus2006critical}
\bibfield{author}{\bibinfo{person}{Derek~C Angus}, \bibinfo{person}{Andrew~F
  Shorr}, \bibinfo{person}{Alan White}, \bibinfo{person}{Tony~T Dremsizov},
  \bibinfo{person}{Robert~J Schmitz}, \bibinfo{person}{Mark~A Kelley},
  \bibinfo{person}{Committee on~Manpower~for Pulmonary},
  \bibinfo{person}{Critical Care~Societies (COMPACCS}, {et~al\mbox{.}}}
  \bibinfo{year}{2006}\natexlab{}.
\newblock \showarticletitle{Critical care delivery in the United States:
  distribution of services and compliance with Leapfrog recommendations}.
\newblock \bibinfo{journal}{\emph{Critical care medicine}}
  \bibinfo{volume}{34}, \bibinfo{number}{4} (\bibinfo{year}{2006}),
  \bibinfo{pages}{1016--1024}.
\newblock


\bibitem[Artsi et~al\mbox{.}(2024)]%
        {artsi2024large}
\bibfield{author}{\bibinfo{person}{Yaara Artsi}, \bibinfo{person}{Vera Sorin},
  \bibinfo{person}{Eli Konen}, \bibinfo{person}{Benjamin~S Glicksberg},
  \bibinfo{person}{Girish Nadkarni}, {and} \bibinfo{person}{Eyal Klang}.}
  \bibinfo{year}{2024}\natexlab{}.
\newblock \showarticletitle{Large language models in simplifying radiological
  reports: systematic review}.
\newblock \bibinfo{journal}{\emph{medRxiv}} (\bibinfo{year}{2024}),
  \bibinfo{pages}{2024--01}.
\newblock


\bibitem[Au et~al\mbox{.}(2017)]%
        {au2017family}
\bibfield{author}{\bibinfo{person}{Selena~S Au}, \bibinfo{person}{Amanda~Roze
  des Ordons}, \bibinfo{person}{Andrea Soo}, \bibinfo{person}{Simon
  Guienguere}, {and} \bibinfo{person}{Henry~T Stelfox}.}
  \bibinfo{year}{2017}\natexlab{}.
\newblock \showarticletitle{Family participation in intensive care unit rounds:
  Comparing family and provider perspectives}.
\newblock \bibinfo{journal}{\emph{Journal of Critical Care}}
  \bibinfo{volume}{38} (\bibinfo{year}{2017}), \bibinfo{pages}{132--136}.
\newblock


\bibitem[Azoulay et~al\mbox{.}(2016)]%
        {azoulay2016communication}
\bibfield{author}{\bibinfo{person}{Elie Azoulay}, \bibinfo{person}{Nancy
  Kentish-Barnes}, {and} \bibinfo{person}{Judith~E Nelson}.}
  \bibinfo{year}{2016}\natexlab{}.
\newblock \showarticletitle{Communication with family caregivers in the
  intensive care unit: Answers and questions}.
\newblock \bibinfo{journal}{\emph{JAMA}} \bibinfo{volume}{315},
  \bibinfo{number}{19} (\bibinfo{year}{2016}), \bibinfo{pages}{2075--2077}.
\newblock


\bibitem[Azoulay et~al\mbox{.}(2002)]%
        {azoulay2002impact}
\bibfield{author}{\bibinfo{person}{Elie Azoulay}, \bibinfo{person}{Frederic
  Pochard}, \bibinfo{person}{Sylvie Chevret}, \bibinfo{person}{Merce Jourdain},
  \bibinfo{person}{Caroline Bornstain}, \bibinfo{person}{Anne Wernet},
  \bibinfo{person}{Isabelle Cattaneo}, \bibinfo{person}{Djilali Annane},
  \bibinfo{person}{Fr{\'e}d{\'e}ric Brun}, \bibinfo{person}{Pierre-Edouard
  Bollaert}, {et~al\mbox{.}}} \bibinfo{year}{2002}\natexlab{}.
\newblock \showarticletitle{Impact of a family information leaflet on
  effectiveness of information provided to family members of intensive care
  unit patients: a multicenter, prospective, randomized, controlled trial}.
\newblock \bibinfo{journal}{\emph{American journal of respiratory and critical
  care medicine}} \bibinfo{volume}{165}, \bibinfo{number}{4}
  (\bibinfo{year}{2002}), \bibinfo{pages}{438--442}.
\newblock


\bibitem[Baker(2001)]%
        {baker2001crossing}
\bibfield{author}{\bibinfo{person}{Alastair Baker}.}
  \bibinfo{year}{2001}\natexlab{}.
\newblock \bibinfo{booktitle}{\emph{Crossing the quality chasm: a new health
  system for the 21st century}}. Vol.~\bibinfo{volume}{323}.
\newblock \bibinfo{publisher}{British Medical Journal Publishing Group}.
\newblock


\bibitem[Barbarin et~al\mbox{.}(2015)]%
        {Barbarin2015}
\bibfield{author}{\bibinfo{person}{Andrea Barbarin},
  \bibinfo{person}{Tiffany~C. Veinot}, {and} \bibinfo{person}{Predrag
  Klasnja}.} \bibinfo{year}{2015}\natexlab{}.
\newblock \showarticletitle{Taking our Time: Chronic Illness and Time-Based
  Objects in Families}. In \bibinfo{booktitle}{\emph{Proceedings of the 18th
  ACM Conference on Computer Supported Cooperative Work \& Social Computing}}
  (Vancouver, BC, Canada) \emph{(\bibinfo{series}{CSCW '15})}.
  \bibinfo{publisher}{Association for Computing Machinery},
  \bibinfo{address}{New York, NY, USA}, \bibinfo{pages}{288–301}.
\newblock
\showISBNx{9781450329224}
\urldef\tempurl%
\url{https://doi.org/10.1145/2675133.2675200}
\showDOI{\tempurl}


\bibitem[Barbarossa et~al\mbox{.}(2023)]%
        {Barbarossa2023}
\bibfield{author}{\bibinfo{person}{Federico Barbarossa},
  \bibinfo{person}{Giulio Amabili}, \bibinfo{person}{Arianna Margaritini},
  \bibinfo{person}{Nicole Morresi}, \bibinfo{person}{Sara Casaccia},
  \bibinfo{person}{Fabrizio Marconi}, \bibinfo{person}{Yeh-Liang Hsu},
  \bibinfo{person}{Fong-Chin Su}, \bibinfo{person}{Nathalie Stolwijk},
  \bibinfo{person}{Henk~Herman Nap}, \bibinfo{person}{Elvira Maranesi}, {and}
  \bibinfo{person}{Roberta Bevlacqua}.} \bibinfo{year}{2023}\natexlab{}.
\newblock \showarticletitle{Design, development, and usability evaluation of a
  dashboard for supporting formal caregivers in managing people with
  dementia.}. In \bibinfo{booktitle}{\emph{Proceedings of the 16th
  International Conference on PErvasive Technologies Related to Assistive
  Environments}} (Corfu, Greece) \emph{(\bibinfo{series}{PETRA '23})}.
  \bibinfo{publisher}{Association for Computing Machinery},
  \bibinfo{address}{New York, NY, USA}, \bibinfo{pages}{154–161}.
\newblock
\showISBNx{9798400700699}
\urldef\tempurl%
\url{https://doi.org/10.1145/3594806.3594820}
\showDOI{\tempurl}


\bibitem[Basu et~al\mbox{.}(2023)]%
        {basu2023med}
\bibfield{author}{\bibinfo{person}{Chandrayee Basu}, \bibinfo{person}{Rosni
  Vasu}, \bibinfo{person}{Michihiro Yasunaga}, {and} \bibinfo{person}{Qian
  Yang}.} \bibinfo{year}{2023}\natexlab{}.
\newblock \showarticletitle{Med-easi: Finely annotated dataset and models for
  controllable simplification of medical texts}. In
  \bibinfo{booktitle}{\emph{Proceedings of the AAAI Conference on Artificial
  Intelligence}}, Vol.~\bibinfo{volume}{37}. \bibinfo{pages}{14093--14101}.
\newblock


\bibitem[Bhat et~al\mbox{.}(2023)]%
        {Bhat2023}
\bibfield{author}{\bibinfo{person}{Karthik~S. Bhat}, \bibinfo{person}{Amanda~K.
  Hall}, \bibinfo{person}{Tiffany Kuo}, {and} \bibinfo{person}{Neha Kumar}.}
  \bibinfo{year}{2023}\natexlab{}.
\newblock \showarticletitle{"We are half-doctors": Family Caregivers as
  Boundary Actors in Chronic Disease Management}.
\newblock \bibinfo{journal}{\emph{Proc. ACM Hum.-Comput. Interact.}}
  \bibinfo{volume}{7}, \bibinfo{number}{CSCW1}, Article
  \bibinfo{articleno}{111} (\bibinfo{date}{April} \bibinfo{year}{2023}),
  \bibinfo{numpages}{29}~pages.
\newblock
\urldef\tempurl%
\url{https://doi.org/10.1145/3579545}
\showDOI{\tempurl}


\bibitem[Bowers et~al\mbox{.}(2024)]%
        {Bowers2024}
\bibfield{author}{\bibinfo{person}{Christopher Bowers}, \bibinfo{person}{Andrew
  Tomlinson}, \bibinfo{person}{Kerry~L Gaskin}, {and} \bibinfo{person}{Jo
  Wray}.} \bibinfo{year}{2024}\natexlab{}.
\newblock \showarticletitle{CHAT2App: Supporting Caregivers of Infants with
  Congenital Heart Disease}. In \bibinfo{booktitle}{\emph{Extended Abstracts of
  the CHI Conference on Human Factors in Computing Systems}} (Honolulu, HI,
  USA) \emph{(\bibinfo{series}{CHI EA '24})}. \bibinfo{publisher}{Association
  for Computing Machinery}, \bibinfo{address}{New York, NY, USA}, Article
  \bibinfo{articleno}{508}, \bibinfo{numpages}{9}~pages.
\newblock
\showISBNx{9798400703317}
\urldef\tempurl%
\url{https://doi.org/10.1145/3613905.3637450}
\showDOI{\tempurl}


\bibitem[Breslow and Stone(2005)]%
        {breslow2005technology}
\bibfield{author}{\bibinfo{person}{Michael~J Breslow} {and}
  \bibinfo{person}{David~J Stone}.} \bibinfo{year}{2005}\natexlab{}.
\newblock \showarticletitle{Technology strategies to improve ICU practice}. In
  \bibinfo{booktitle}{\emph{Seminars in Anesthesia, Perioperative Medicine and
  Pain}}, Vol.~\bibinfo{volume}{24}. Elsevier, \bibinfo{pages}{59--70}.
\newblock


\bibitem[Caligtan et~al\mbox{.}(2012)]%
        {caligtan2012bedside}
\bibfield{author}{\bibinfo{person}{Christine~A Caligtan},
  \bibinfo{person}{Diane~L Carroll}, \bibinfo{person}{Ann~C Hurley},
  \bibinfo{person}{Ronna Gersh-Zaremski}, {and} \bibinfo{person}{Patricia~C
  Dykes}.} \bibinfo{year}{2012}\natexlab{}.
\newblock \showarticletitle{Bedside information technology to support
  patient-centered care}.
\newblock \bibinfo{journal}{\emph{International journal of medical
  informatics}} \bibinfo{volume}{81}, \bibinfo{number}{7}
  (\bibinfo{year}{2012}), \bibinfo{pages}{442--451}.
\newblock


\bibitem[CARE et~al\mbox{.}(2012)]%
        {care2012patient}
\bibfield{author}{\bibinfo{person}{COMMITTEE ON~HOSPITAL CARE},
  \bibinfo{person}{INSTITUTE~FOR PATIENT}, {and}
  \bibinfo{person}{FAMILY-CENTERED CARE}.} \bibinfo{year}{2012}\natexlab{}.
\newblock \showarticletitle{Patient-and family-centered care and the
  pediatrician's role}.
\newblock \bibinfo{journal}{\emph{Pediatrics}} \bibinfo{volume}{129},
  \bibinfo{number}{2} (\bibinfo{year}{2012}), \bibinfo{pages}{394--404}.
\newblock


\bibitem[Chen et~al\mbox{.}(2013)]%
        {Yunan2013}
\bibfield{author}{\bibinfo{person}{Yunan Chen}, \bibinfo{person}{Victor Ngo},
  {and} \bibinfo{person}{Sun~Young Park}.} \bibinfo{year}{2013}\natexlab{}.
\newblock \showarticletitle{Caring for caregivers: designing for integrality}.
  In \bibinfo{booktitle}{\emph{Proceedings of the 2013 Conference on Computer
  Supported Cooperative Work}} (San Antonio, Texas, USA)
  \emph{(\bibinfo{series}{CSCW '13})}. \bibinfo{publisher}{Association for
  Computing Machinery}, \bibinfo{address}{New York, NY, USA},
  \bibinfo{pages}{91–102}.
\newblock
\showISBNx{9781450313315}
\urldef\tempurl%
\url{https://doi.org/10.1145/2441776.2441789}
\showDOI{\tempurl}


\bibitem[Choi et~al\mbox{.}(2019)]%
        {choi2019exploring}
\bibfield{author}{\bibinfo{person}{JiYeon Choi}, \bibinfo{person}{Youn-Jung
  Son}, {and} \bibinfo{person}{Judith~A Tate}.}
  \bibinfo{year}{2019}\natexlab{}.
\newblock \showarticletitle{Exploring positive aspects of caregiving in family
  caregivers of adult ICU survivors from ICU to four months post-ICU
  discharge}.
\newblock \bibinfo{journal}{\emph{Heart \& Lung}} \bibinfo{volume}{48},
  \bibinfo{number}{6} (\bibinfo{year}{2019}), \bibinfo{pages}{553--559}.
\newblock


\bibitem[Choi et~al\mbox{.}(2014)]%
        {choi2014fatigue}
\bibfield{author}{\bibinfo{person}{JiYeon Choi}, \bibinfo{person}{Judith~A
  Tate}, \bibinfo{person}{Leslie~A Hoffman}, \bibinfo{person}{Richard Schulz},
  \bibinfo{person}{Dianxu Ren}, \bibinfo{person}{Michael~P Donahoe},
  \bibinfo{person}{Barbara~A Given}, {and} \bibinfo{person}{Paula~R Sherwood}.}
  \bibinfo{year}{2014}\natexlab{}.
\newblock \showarticletitle{Fatigue in family caregivers of adult intensive
  care unit survivors}.
\newblock \bibinfo{journal}{\emph{Journal of pain and symptom management}}
  \bibinfo{volume}{48}, \bibinfo{number}{3} (\bibinfo{year}{2014}),
  \bibinfo{pages}{353--363}.
\newblock


\bibitem[Corbin and Strauss(2014)]%
        {corbin2014basics}
\bibfield{author}{\bibinfo{person}{Juliet Corbin} {and} \bibinfo{person}{Anselm
  Strauss}.} \bibinfo{year}{2014}\natexlab{}.
\newblock \bibinfo{booktitle}{\emph{Basics of qualitative research: Techniques
  and procedures for developing grounded theory}}.
\newblock \bibinfo{publisher}{Sage publications}.
\newblock


\bibitem[Corey et~al\mbox{.}(2018)]%
        {corey2018development}
\bibfield{author}{\bibinfo{person}{Kristin~M Corey}, \bibinfo{person}{Sehj
  Kashyap}, \bibinfo{person}{Elizabeth Lorenzi}, \bibinfo{person}{Sandhya~A
  Lagoo-Deenadayalan}, \bibinfo{person}{Katherine Heller},
  \bibinfo{person}{Krista Whalen}, \bibinfo{person}{Suresh Balu},
  \bibinfo{person}{Mitchell~T Heflin}, \bibinfo{person}{Shelley~R McDonald},
  \bibinfo{person}{Madhav Swaminathan}, {et~al\mbox{.}}}
  \bibinfo{year}{2018}\natexlab{}.
\newblock \showarticletitle{Development and validation of machine learning
  models to identify high-risk surgical patients using automatically curated
  electronic health record data (Pythia): a retrospective, single-site study}.
\newblock \bibinfo{journal}{\emph{PLoS medicine}} \bibinfo{volume}{15},
  \bibinfo{number}{11} (\bibinfo{year}{2018}), \bibinfo{pages}{e1002701}.
\newblock


\bibitem[Davidson et~al\mbox{.}(2017)]%
        {davidson2017guidelines}
\bibfield{author}{\bibinfo{person}{Judy~E Davidson}, \bibinfo{person}{Rebecca~A
  Aslakson}, \bibinfo{person}{Ann~C Long}, \bibinfo{person}{Kathleen~A
  Puntillo}, \bibinfo{person}{Erin~K Kross}, \bibinfo{person}{Joanna Hart},
  \bibinfo{person}{Christopher~E Cox}, \bibinfo{person}{Hannah Wunsch},
  \bibinfo{person}{Mary~A Wickline}, \bibinfo{person}{Mark~E Nunnally},
  {et~al\mbox{.}}} \bibinfo{year}{2017}\natexlab{}.
\newblock \showarticletitle{Guidelines for family-centered care in the
  neonatal, pediatric, and adult ICU}.
\newblock \bibinfo{journal}{\emph{Critical care medicine}}
  \bibinfo{volume}{45}, \bibinfo{number}{1} (\bibinfo{year}{2017}),
  \bibinfo{pages}{103--128}.
\newblock


\bibitem[Davis et~al\mbox{.}(2005)]%
        {davis20052020}
\bibfield{author}{\bibinfo{person}{Karen Davis}, \bibinfo{person}{Stephen~C
  Schoenbaum}, {and} \bibinfo{person}{Anne-Marie Audet}.}
  \bibinfo{year}{2005}\natexlab{}.
\newblock \showarticletitle{A 2020 vision of patient-centered primary care}.
\newblock \bibinfo{journal}{\emph{Journal of general internal medicine}}
  \bibinfo{volume}{20} (\bibinfo{year}{2005}), \bibinfo{pages}{953--957}.
\newblock


\bibitem[Del~Sette et~al\mbox{.}(2023)]%
        {Sette2023}
\bibfield{author}{\bibinfo{person}{Bleiz~Macsen Del~Sette},
  \bibinfo{person}{Dawn Carnes}, {and} \bibinfo{person}{Charalampos Saitis}.}
  \bibinfo{year}{2023}\natexlab{}.
\newblock \showarticletitle{Sound of Care: Towards a Co-Operative AI Digital
  Pain Companion to Support People with Chronic Primary Pain}. In
  \bibinfo{booktitle}{\emph{Companion Publication of the 2023 Conference on
  Computer Supported Cooperative Work and Social Computing}} (Minneapolis, MN,
  USA) \emph{(\bibinfo{series}{CSCW '23 Companion})}.
  \bibinfo{publisher}{Association for Computing Machinery},
  \bibinfo{address}{New York, NY, USA}, \bibinfo{pages}{283–288}.
\newblock
\showISBNx{9798400701290}
\urldef\tempurl%
\url{https://doi.org/10.1145/3584931.3606971}
\showDOI{\tempurl}


\bibitem[Dragoi et~al\mbox{.}(2022)]%
        {dragoi2022visitation}
\bibfield{author}{\bibinfo{person}{Laura Dragoi}, \bibinfo{person}{Laveena
  Munshi}, {and} \bibinfo{person}{Margaret Herridge}.}
  \bibinfo{year}{2022}\natexlab{}.
\newblock \showarticletitle{Visitation policies in the ICU and the importance
  of family presence at the bedside}.
\newblock \bibinfo{journal}{\emph{Intensive Care Medicine}}
  \bibinfo{volume}{48}, \bibinfo{number}{12} (\bibinfo{year}{2022}),
  \bibinfo{pages}{1790--1792}.
\newblock


\bibitem[Dunst and Trivette(1996)]%
        {dunst1996empowerment}
\bibfield{author}{\bibinfo{person}{Carl~J Dunst} {and} \bibinfo{person}{Carol~M
  Trivette}.} \bibinfo{year}{1996}\natexlab{}.
\newblock \showarticletitle{Empowerment, effective helpgiving practices and
  family-centered care}.
\newblock \bibinfo{journal}{\emph{Pediatric nursing}} \bibinfo{volume}{22},
  \bibinfo{number}{4} (\bibinfo{year}{1996}), \bibinfo{pages}{334--338}.
\newblock


\bibitem[Essen(2004)]%
        {essen2004proxy}
\bibfield{author}{\bibinfo{person}{Louise~von Essen}.}
  \bibinfo{year}{2004}\natexlab{}.
\newblock \showarticletitle{Proxy ratings of patient quality of life Factors
  related to patient--proxy agreement}.
\newblock \bibinfo{journal}{\emph{Acta oncologica}} \bibinfo{volume}{43},
  \bibinfo{number}{3} (\bibinfo{year}{2004}), \bibinfo{pages}{229--234}.
\newblock


\bibitem[Etikan et~al\mbox{.}(2016)]%
        {etikan2016comparison}
\bibfield{author}{\bibinfo{person}{Ilker Etikan},
  \bibinfo{person}{Sulaiman~Abubakar Musa}, \bibinfo{person}{Rukayya~Sunusi
  Alkassim}, {et~al\mbox{.}}} \bibinfo{year}{2016}\natexlab{}.
\newblock \showarticletitle{Comparison of convenience sampling and purposive
  sampling}.
\newblock \bibinfo{journal}{\emph{American journal of theoretical and applied
  statistics}} \bibinfo{volume}{5}, \bibinfo{number}{1} (\bibinfo{year}{2016}),
  \bibinfo{pages}{1--4}.
\newblock


\bibitem[Flaatten et~al\mbox{.}(2017)]%
        {flaatten2017status}
\bibfield{author}{\bibinfo{person}{H Flaatten}, \bibinfo{person}{DW De~Lange},
  \bibinfo{person}{A Artigas}, \bibinfo{person}{D Bin}, \bibinfo{person}{R
  Moreno}, \bibinfo{person}{S Christensen}, \bibinfo{person}{GM Joynt},
  \bibinfo{person}{Sean~M Bagshaw}, \bibinfo{person}{CL Sprung},
  \bibinfo{person}{Dominique Benoit}, {et~al\mbox{.}}}
  \bibinfo{year}{2017}\natexlab{}.
\newblock \showarticletitle{The status of intensive care medicine research and
  a future agenda for very old patients in the ICU}.
\newblock \bibinfo{journal}{\emph{Intensive care medicine}}
  \bibinfo{volume}{43} (\bibinfo{year}{2017}), \bibinfo{pages}{1319--1328}.
\newblock


\bibitem[Foong et~al\mbox{.}(2020)]%
        {Foong2020}
\bibfield{author}{\bibinfo{person}{Pin~Sym Foong}, \bibinfo{person}{Charis~Anne
  Lim}, \bibinfo{person}{Joshua Wong}, \bibinfo{person}{Chang~Siang Lim},
  \bibinfo{person}{Simon~Tangi Perrault}, {and} \bibinfo{person}{Gerald~CH
  Koh}.} \bibinfo{year}{2020}\natexlab{}.
\newblock \showarticletitle{"You Cannot Offer Such a Suggestion": Designing for
  Family Caregiver Input in Home Care Systems}. In
  \bibinfo{booktitle}{\emph{Proceedings of the 2020 CHI Conference on Human
  Factors in Computing Systems}} (Honolulu, HI, USA)
  \emph{(\bibinfo{series}{CHI '20})}. \bibinfo{publisher}{Association for
  Computing Machinery}, \bibinfo{address}{New York, NY, USA},
  \bibinfo{pages}{1–13}.
\newblock
\showISBNx{9781450367080}
\urldef\tempurl%
\url{https://doi.org/10.1145/3313831.3376607}
\showDOI{\tempurl}


\bibitem[Gaeeni et~al\mbox{.}(2014)]%
        {gaeeni2014informational}
\bibfield{author}{\bibinfo{person}{Mina Gaeeni}, \bibinfo{person}{Mansoureh~A
  Farahani}, \bibinfo{person}{Naima Seyedfatemi}, {and}
  \bibinfo{person}{Nooredin Mohammadi}.} \bibinfo{year}{2014}\natexlab{}.
\newblock \showarticletitle{Informational support to family members of
  intensive care unit patients: the perspectives of families and nurses}.
\newblock \bibinfo{journal}{\emph{Global journal of health science}}
  \bibinfo{volume}{7}, \bibinfo{number}{2} (\bibinfo{year}{2014}),
  \bibinfo{pages}{8}.
\newblock


\bibitem[Gao et~al\mbox{.}(2023)]%
        {gao2023retrieval}
\bibfield{author}{\bibinfo{person}{Yunfan Gao}, \bibinfo{person}{Yun Xiong},
  \bibinfo{person}{Xinyu Gao}, \bibinfo{person}{Kangxiang Jia},
  \bibinfo{person}{Jinliu Pan}, \bibinfo{person}{Yuxi Bi},
  \bibinfo{person}{Yixin Dai}, \bibinfo{person}{Jiawei Sun},
  \bibinfo{person}{Haofen Wang}, {and} \bibinfo{person}{Haofen Wang}.}
  \bibinfo{year}{2023}\natexlab{}.
\newblock \showarticletitle{Retrieval-augmented generation for large language
  models: A survey}.
\newblock \bibinfo{journal}{\emph{arXiv preprint arXiv:2312.10997}}
  \bibinfo{volume}{2}, \bibinfo{number}{1} (\bibinfo{year}{2023}).
\newblock


\bibitem[Goldfarb et~al\mbox{.}(2017)]%
        {goldfarb2017outcomes}
\bibfield{author}{\bibinfo{person}{Michael~J Goldfarb}, \bibinfo{person}{Lior
  Bibas}, \bibinfo{person}{Virginia Bartlett}, \bibinfo{person}{Heather Jones},
  {and} \bibinfo{person}{Naureen Khan}.} \bibinfo{year}{2017}\natexlab{}.
\newblock \showarticletitle{Outcomes of patient-and family-centered care
  interventions in the ICU: a systematic review and meta-analysis}.
\newblock \bibinfo{journal}{\emph{Critical care medicine}}
  \bibinfo{volume}{45}, \bibinfo{number}{10} (\bibinfo{year}{2017}),
  \bibinfo{pages}{1751--1761}.
\newblock


\bibitem[Gon{\c{c}}alves et~al\mbox{.}(2023)]%
        {gonccalves2023surviving}
\bibfield{author}{\bibinfo{person}{Ana-Carolina Gon{\c{c}}alves},
  \bibinfo{person}{Annabel Williams}, \bibinfo{person}{Christina Koulouglioti},
  \bibinfo{person}{Todd Leckie}, \bibinfo{person}{Alexander Hunter},
  \bibinfo{person}{Daniel Fitzpatrick}, \bibinfo{person}{Alan Richardson},
  \bibinfo{person}{Benjamin Hardy}, \bibinfo{person}{Richard Venn}, {and}
  \bibinfo{person}{Luke Hodgson}.} \bibinfo{year}{2023}\natexlab{}.
\newblock \showarticletitle{Surviving severe COVID-19: Interviews with
  patients, informal carers and health professionals}.
\newblock \bibinfo{journal}{\emph{Nursing in critical care}}
  \bibinfo{volume}{28}, \bibinfo{number}{1} (\bibinfo{year}{2023}),
  \bibinfo{pages}{80--88}.
\newblock


\bibitem[Gopinath et~al\mbox{.}(2020)]%
        {gopinath2020fast}
\bibfield{author}{\bibinfo{person}{Divya Gopinath}, \bibinfo{person}{Monica
  Agrawal}, \bibinfo{person}{Luke Murray}, \bibinfo{person}{Steven Horng},
  \bibinfo{person}{David Karger}, {and} \bibinfo{person}{David Sontag}.}
  \bibinfo{year}{2020}\natexlab{}.
\newblock \showarticletitle{Fast, structured clinical documentation via
  contextual autocomplete}. In \bibinfo{booktitle}{\emph{Machine Learning for
  Healthcare Conference}}. PMLR, \bibinfo{pages}{842--870}.
\newblock


\bibitem[Grieve et~al\mbox{.}(2019)]%
        {grieve2019analysis}
\bibfield{author}{\bibinfo{person}{Richard Grieve}, \bibinfo{person}{Stephen
  O’Neill}, \bibinfo{person}{Anirban Basu}, \bibinfo{person}{Luke Keele},
  \bibinfo{person}{Kathryn~M Rowan}, {and} \bibinfo{person}{Steve Harris}.}
  \bibinfo{year}{2019}\natexlab{}.
\newblock \showarticletitle{Analysis of benefit of intensive care unit transfer
  for deteriorating ward patients: a patient-centered approach to clinical
  evaluation}.
\newblock \bibinfo{journal}{\emph{JAMA network open}} \bibinfo{volume}{2},
  \bibinfo{number}{2} (\bibinfo{year}{2019}),
  \bibinfo{pages}{e187704--e187704}.
\newblock


\bibitem[Hanberger et~al\mbox{.}(2005)]%
        {hanberger2005intensive}
\bibfield{author}{\bibinfo{person}{Hakan Hanberger},
  \bibinfo{person}{Dominique~L Monnet}, {and} \bibinfo{person}{Lennart~E
  Nilsson}.} \bibinfo{year}{2005}\natexlab{}.
\newblock \showarticletitle{Intensive care unit}.
\newblock In \bibinfo{booktitle}{\emph{Antibiotic policies: theory and
  practice}}. \bibinfo{publisher}{Springer}, \bibinfo{pages}{261--279}.
\newblock


\bibitem[Hao et~al\mbox{.}(2024)]%
        {Yuexing2024}
\bibfield{author}{\bibinfo{person}{Yuexing Hao}, \bibinfo{person}{Corinna~E.
  L\"{o}eckenhoff}, \bibinfo{person}{Harry Lee}, \bibinfo{person}{Jessica
  Zwerling}, {and} \bibinfo{person}{Saleh Kalantari}.}
  \bibinfo{year}{2024}\natexlab{}.
\newblock \showarticletitle{The i-SDM Framework: Developing AI-based Tools in
  Shared Decision-Making for Cancer Treatment with Clinical Professionals}. In
  \bibinfo{booktitle}{\emph{Companion Publication of the 2024 Conference on
  Computer-Supported Cooperative Work and Social Computing}} (San Jose, Costa
  Rica) \emph{(\bibinfo{series}{CSCW Companion '24})}.
  \bibinfo{publisher}{Association for Computing Machinery},
  \bibinfo{address}{New York, NY, USA}, \bibinfo{pages}{134–140}.
\newblock
\showISBNx{9798400711145}
\urldef\tempurl%
\url{https://doi.org/10.1145/3678884.3681841}
\showDOI{\tempurl}


\bibitem[H{\"a}yrinen et~al\mbox{.}(2008)]%
        {hayrinen2008definition}
\bibfield{author}{\bibinfo{person}{Kristiina H{\"a}yrinen},
  \bibinfo{person}{Kaija Saranto}, {and} \bibinfo{person}{Pirkko Nyk{\"a}nen}.}
  \bibinfo{year}{2008}\natexlab{}.
\newblock \showarticletitle{Definition, structure, content, use and impacts of
  electronic health records: a review of the research literature}.
\newblock \bibinfo{journal}{\emph{International journal of medical
  informatics}} \bibinfo{volume}{77}, \bibinfo{number}{5}
  (\bibinfo{year}{2008}), \bibinfo{pages}{291--304}.
\newblock


\bibitem[Hines et~al\mbox{.}(2011)]%
        {hines2011effectiveness}
\bibfield{author}{\bibinfo{person}{Sonia Hines}, \bibinfo{person}{Judy McCrow},
  \bibinfo{person}{Jenny Abbey}, \bibinfo{person}{Jenneke Foottit},
  \bibinfo{person}{Jacinda Wilson}, \bibinfo{person}{Sara Franklin}, {and}
  \bibinfo{person}{Elizabeth Beattie}.} \bibinfo{year}{2011}\natexlab{}.
\newblock \showarticletitle{The effectiveness and appropriateness of a
  palliative approach to care for people with advanced dementia: a systematic
  review}.
\newblock \bibinfo{journal}{\emph{JBI Evidence Synthesis}} \bibinfo{volume}{9},
  \bibinfo{number}{26} (\bibinfo{year}{2011}), \bibinfo{pages}{960--1131}.
\newblock


\bibitem[Hong et~al\mbox{.}(2017)]%
        {Hong2017}
\bibfield{author}{\bibinfo{person}{Matthew~K. Hong}, \bibinfo{person}{Clayton
  Feustel}, \bibinfo{person}{Meeshu Agnihotri}, \bibinfo{person}{Max
  Silverman}, \bibinfo{person}{Stephen~F. Simoneaux}, {and}
  \bibinfo{person}{Lauren Wilcox}.} \bibinfo{year}{2017}\natexlab{}.
\newblock \showarticletitle{Supporting Families in Reviewing and Communicating
  about Radiology Imaging Studies}. In \bibinfo{booktitle}{\emph{Proceedings of
  the 2017 CHI Conference on Human Factors in Computing Systems}} (Denver,
  Colorado, USA) \emph{(\bibinfo{series}{CHI '17})}.
  \bibinfo{publisher}{Association for Computing Machinery},
  \bibinfo{address}{New York, NY, USA}, \bibinfo{pages}{5245–5256}.
\newblock
\showISBNx{9781450346559}
\urldef\tempurl%
\url{https://doi.org/10.1145/3025453.3025754}
\showDOI{\tempurl}


\bibitem[Huo et~al\mbox{.}(2025)]%
        {huo2025large}
\bibfield{author}{\bibinfo{person}{Bright Huo}, \bibinfo{person}{Amy Boyle},
  \bibinfo{person}{Nana Marfo}, \bibinfo{person}{Wimonchat Tangamornsuksan},
  \bibinfo{person}{Jeremy~P Steen}, \bibinfo{person}{Tyler McKechnie},
  \bibinfo{person}{Yung Lee}, \bibinfo{person}{Julio Mayol},
  \bibinfo{person}{Stavros~A Antoniou}, \bibinfo{person}{Arun~James
  Thirunavukarasu}, {et~al\mbox{.}}} \bibinfo{year}{2025}\natexlab{}.
\newblock \showarticletitle{Large Language Models for Chatbot Health Advice
  Studies: A Systematic Review}.
\newblock \bibinfo{journal}{\emph{JAMA Network Open}} \bibinfo{volume}{8},
  \bibinfo{number}{2} (\bibinfo{year}{2025}),
  \bibinfo{pages}{e2457879--e2457879}.
\newblock


\bibitem[Hwang et~al\mbox{.}(2014)]%
        {Hwang2014}
\bibfield{author}{\bibinfo{person}{Inseok Hwang}, \bibinfo{person}{Chungkuk
  Yoo}, \bibinfo{person}{Chanyou Hwang}, \bibinfo{person}{Dongsun Yim},
  \bibinfo{person}{Youngki Lee}, \bibinfo{person}{Chulhong Min},
  \bibinfo{person}{John Kim}, {and} \bibinfo{person}{Junehwa Song}.}
  \bibinfo{year}{2014}\natexlab{}.
\newblock \showarticletitle{TalkBetter: family-driven mobile intervention care
  for children with language delay}. In \bibinfo{booktitle}{\emph{Proceedings
  of the 17th ACM Conference on Computer Supported Cooperative Work \& Social
  Computing}} (Baltimore, Maryland, USA) \emph{(\bibinfo{series}{CSCW '14})}.
  \bibinfo{publisher}{Association for Computing Machinery},
  \bibinfo{address}{New York, NY, USA}, \bibinfo{pages}{1283–1296}.
\newblock
\showISBNx{9781450325400}
\urldef\tempurl%
\url{https://doi.org/10.1145/2531602.2531668}
\showDOI{\tempurl}


\bibitem[Jagannatha et~al\mbox{.}(2019)]%
        {jagannatha2019overview}
\bibfield{author}{\bibinfo{person}{Abhyuday Jagannatha},
  \bibinfo{person}{Feifan Liu}, \bibinfo{person}{Weisong Liu}, {and}
  \bibinfo{person}{Hong Yu}.} \bibinfo{year}{2019}\natexlab{}.
\newblock \showarticletitle{Overview of the first natural language processing
  challenge for extracting medication, indication, and adverse drug events from
  electronic health record notes (MADE 1.0)}.
\newblock \bibinfo{journal}{\emph{Drug safety}} \bibinfo{volume}{42},
  \bibinfo{number}{1} (\bibinfo{year}{2019}), \bibinfo{pages}{99--111}.
\newblock


\bibitem[Jarvis(2023)]%
        {jarvis2023physical}
\bibfield{author}{\bibinfo{person}{Carolyn Jarvis}.}
  \bibinfo{year}{2023}\natexlab{}.
\newblock \bibinfo{booktitle}{\emph{Physical Examination and Health
  Assessment-Canadian E-Book: Physical Examination and Health
  Assessment-Canadian E-Book}}.
\newblock \bibinfo{publisher}{Elsevier Health Sciences}.
\newblock


\bibitem[Jennerich et~al\mbox{.}(2020)]%
        {jennerich2020unplanned}
\bibfield{author}{\bibinfo{person}{Ann~L Jennerich}, \bibinfo{person}{Mara~R
  Hobler}, \bibinfo{person}{Rashmi~K Sharma}, \bibinfo{person}{Ruth~A
  Engelberg}, {and} \bibinfo{person}{J~Randall Curtis}.}
  \bibinfo{year}{2020}\natexlab{}.
\newblock \showarticletitle{Unplanned admission to the ICU: a qualitative study
  examining family member experiences}.
\newblock \bibinfo{journal}{\emph{Chest}} \bibinfo{volume}{158},
  \bibinfo{number}{4} (\bibinfo{year}{2020}), \bibinfo{pages}{1482--1489}.
\newblock


\bibitem[Jo et~al\mbox{.}(2024)]%
        {Eunkyung2024}
\bibfield{author}{\bibinfo{person}{Eunkyung Jo}, \bibinfo{person}{Rachael
  Zehrung}, \bibinfo{person}{Katherine Genuario}, \bibinfo{person}{Alexandra
  Papoutsaki}, {and} \bibinfo{person}{Daniel~A. Epstein}.}
  \bibinfo{year}{2024}\natexlab{}.
\newblock \showarticletitle{Exploring Patient-Generated Annotations to Digital
  Clinical Symptom Measures for Patient-Centered Communication}.
\newblock \bibinfo{journal}{\emph{Proc. ACM Hum.-Comput. Interact.}}
  \bibinfo{volume}{8}, \bibinfo{number}{CSCW2}, Article
  \bibinfo{articleno}{458} (\bibinfo{date}{Nov.} \bibinfo{year}{2024}),
  \bibinfo{numpages}{26}~pages.
\newblock
\urldef\tempurl%
\url{https://doi.org/10.1145/3686997}
\showDOI{\tempurl}


\bibitem[Johnson et~al\mbox{.}(2016)]%
        {johnson2016mimic}
\bibfield{author}{\bibinfo{person}{Alistair~EW Johnson}, \bibinfo{person}{Tom~J
  Pollard}, \bibinfo{person}{Lu Shen}, \bibinfo{person}{Li-wei~H Lehman},
  \bibinfo{person}{Mengling Feng}, \bibinfo{person}{Mohammad Ghassemi},
  \bibinfo{person}{Benjamin Moody}, \bibinfo{person}{Peter Szolovits},
  \bibinfo{person}{Leo Anthony~Celi}, {and} \bibinfo{person}{Roger~G Mark}.}
  \bibinfo{year}{2016}\natexlab{}.
\newblock \showarticletitle{MIMIC-III, a freely accessible critical care
  database}.
\newblock \bibinfo{journal}{\emph{Scientific data}} \bibinfo{volume}{3},
  \bibinfo{number}{1} (\bibinfo{year}{2016}), \bibinfo{pages}{1--9}.
\newblock


\bibitem[Johnson et~al\mbox{.}(1995)]%
        {johnson1995perceived}
\bibfield{author}{\bibinfo{person}{Susan~K Johnson}, \bibinfo{person}{Martha
  Craft}, \bibinfo{person}{Marita Titler}, \bibinfo{person}{Margo Halm},
  \bibinfo{person}{Charmaine Kleiber}, \bibinfo{person}{Lou~Ann Montgomery},
  \bibinfo{person}{Karen Megivern}, \bibinfo{person}{Anita Nicholson}, {and}
  \bibinfo{person}{Kathleen Buckwalter}.} \bibinfo{year}{1995}\natexlab{}.
\newblock \showarticletitle{Perceived changes in adult family members' roles
  and responsibilities during critical illness}.
\newblock \bibinfo{journal}{\emph{Image: the Journal of Nursing Scholarship}}
  \bibinfo{volume}{27}, \bibinfo{number}{3} (\bibinfo{year}{1995}),
  \bibinfo{pages}{238--243}.
\newblock


\bibitem[Jolley and Shields(2009)]%
        {jolley2009evolution}
\bibfield{author}{\bibinfo{person}{Jeremy Jolley} {and} \bibinfo{person}{Linda
  Shields}.} \bibinfo{year}{2009}\natexlab{}.
\newblock \showarticletitle{The evolution of family-centered care}.
\newblock \bibinfo{journal}{\emph{Journal of pediatric nursing}}
  \bibinfo{volume}{24}, \bibinfo{number}{2} (\bibinfo{year}{2009}),
  \bibinfo{pages}{164--170}.
\newblock


\bibitem[Kaziunas et~al\mbox{.}(2015)]%
        {Kaziunas2015}
\bibfield{author}{\bibinfo{person}{Elizabeth Kaziunas},
  \bibinfo{person}{Ayse~G. Buyuktur}, \bibinfo{person}{Jasmine Jones},
  \bibinfo{person}{Sung~W. Choi}, \bibinfo{person}{David~A. Hanauer}, {and}
  \bibinfo{person}{Mark~S. Ackerman}.} \bibinfo{year}{2015}\natexlab{}.
\newblock \showarticletitle{Transition and Reflection in the Use of Health
  Information: The Case of Pediatric Bone Marrow Transplant Caregivers}. In
  \bibinfo{booktitle}{\emph{Proceedings of the 18th ACM Conference on Computer
  Supported Cooperative Work \& Social Computing}} (Vancouver, BC, Canada)
  \emph{(\bibinfo{series}{CSCW '15})}. \bibinfo{publisher}{Association for
  Computing Machinery}, \bibinfo{address}{New York, NY, USA},
  \bibinfo{pages}{1763–1774}.
\newblock
\showISBNx{9781450329224}
\urldef\tempurl%
\url{https://doi.org/10.1145/2675133.2675276}
\showDOI{\tempurl}


\bibitem[Kong et~al\mbox{.}(2017)]%
        {kong2017comparative}
\bibfield{author}{\bibinfo{person}{Ha~Kyung Kong}, \bibinfo{person}{John Lee},
  {and} \bibinfo{person}{Karrie Karahalios}.} \bibinfo{year}{2017}\natexlab{}.
\newblock \showarticletitle{A comparative study of visualizations with
  different granularities of behavior for communicating about autism}.
\newblock \bibinfo{journal}{\emph{Proceedings of the ACM on Human-Computer
  Interaction}} \bibinfo{volume}{1}, \bibinfo{number}{CSCW}
  (\bibinfo{year}{2017}), \bibinfo{pages}{1--16}.
\newblock


\bibitem[Kruse et~al\mbox{.}(2025)]%
        {kruse2025zero}
\bibfield{author}{\bibinfo{person}{Maya Kruse}, \bibinfo{person}{Shiyue Hu},
  \bibinfo{person}{Nicholas Derby}, \bibinfo{person}{Yifu Wu},
  \bibinfo{person}{Samantha Stonbraker}, \bibinfo{person}{Bingsheng Yao},
  \bibinfo{person}{Dakuo Wang}, \bibinfo{person}{Elizabeth Goldberg}, {and}
  \bibinfo{person}{Yanjun Gao}.} \bibinfo{year}{2025}\natexlab{}.
\newblock \showarticletitle{Zero-Shot Large Language Models for Long Clinical
  Text Summarization with Temporal Reasoning}.
\newblock \bibinfo{journal}{\emph{arXiv preprint arXiv:2501.18724}}
  (\bibinfo{year}{2025}).
\newblock


\bibitem[Lewis et~al\mbox{.}(2020)]%
        {lewis2020retrieval}
\bibfield{author}{\bibinfo{person}{Patrick Lewis}, \bibinfo{person}{Ethan
  Perez}, \bibinfo{person}{Aleksandra Piktus}, \bibinfo{person}{Fabio Petroni},
  \bibinfo{person}{Vladimir Karpukhin}, \bibinfo{person}{Naman Goyal},
  \bibinfo{person}{Heinrich K{\"u}ttler}, \bibinfo{person}{Mike Lewis},
  \bibinfo{person}{Wen-tau Yih}, \bibinfo{person}{Tim Rockt{\"a}schel},
  {et~al\mbox{.}}} \bibinfo{year}{2020}\natexlab{}.
\newblock \showarticletitle{Retrieval-augmented generation for
  knowledge-intensive nlp tasks}.
\newblock \bibinfo{journal}{\emph{Advances in neural information processing
  systems}}  \bibinfo{volume}{33} (\bibinfo{year}{2020}),
  \bibinfo{pages}{9459--9474}.
\newblock


\bibitem[Liu et~al\mbox{.}(2011)]%
        {liu2011}
\bibfield{author}{\bibinfo{person}{Leslie~S. Liu}, \bibinfo{person}{Sen~H.
  Hirano}, \bibinfo{person}{Monica Tentori}, \bibinfo{person}{Karen~G. Cheng},
  \bibinfo{person}{Sheba George}, \bibinfo{person}{Sun~Young Park}, {and}
  \bibinfo{person}{Gillian~R. Hayes}.} \bibinfo{year}{2011}\natexlab{}.
\newblock \showarticletitle{Improving communication and social support for
  caregivers of high-risk infants through mobile technologies}. In
  \bibinfo{booktitle}{\emph{Proceedings of the ACM 2011 Conference on Computer
  Supported Cooperative Work}} (Hangzhou, China) \emph{(\bibinfo{series}{CSCW
  '11})}. \bibinfo{publisher}{Association for Computing Machinery},
  \bibinfo{address}{New York, NY, USA}, \bibinfo{pages}{475–484}.
\newblock
\showISBNx{9781450305563}
\urldef\tempurl%
\url{https://doi.org/10.1145/1958824.1958897}
\showDOI{\tempurl}


\bibitem[Longhurst(2003)]%
        {longhurst2003semi}
\bibfield{author}{\bibinfo{person}{Robyn Longhurst}.}
  \bibinfo{year}{2003}\natexlab{}.
\newblock \showarticletitle{Semi-structured interviews and focus groups}.
\newblock \bibinfo{journal}{\emph{Key methods in geography}}
  \bibinfo{volume}{3}, \bibinfo{number}{2} (\bibinfo{year}{2003}),
  \bibinfo{pages}{143--156}.
\newblock


\bibitem[McGonigal(2020)]%
        {mcgonigal2020providing}
\bibfield{author}{\bibinfo{person}{Michelle McGonigal}.}
  \bibinfo{year}{2020}\natexlab{}.
\newblock \showarticletitle{Providing quality care to the intellectually
  disadvantaged patient population during the COVID-19 pandemic}.
\newblock \bibinfo{journal}{\emph{Critical care nursing quarterly}}
  \bibinfo{volume}{43}, \bibinfo{number}{4} (\bibinfo{year}{2020}),
  \bibinfo{pages}{480--483}.
\newblock


\bibitem[Meert et~al\mbox{.}(2013)]%
        {meert2013family}
\bibfield{author}{\bibinfo{person}{Kathleen~L Meert}, \bibinfo{person}{Jeff
  Clark}, {and} \bibinfo{person}{Susan Eggly}.}
  \bibinfo{year}{2013}\natexlab{}.
\newblock \showarticletitle{Family-centered care in the pediatric intensive
  care unit}.
\newblock \bibinfo{journal}{\emph{Pediatric Clinics}} \bibinfo{volume}{60},
  \bibinfo{number}{3} (\bibinfo{year}{2013}), \bibinfo{pages}{761--772}.
\newblock


\bibitem[Miller et~al\mbox{.}(2016)]%
        {Miller2016}
\bibfield{author}{\bibinfo{person}{Andrew~D. Miller},
  \bibinfo{person}{Sonali~R. Mishra}, \bibinfo{person}{Logan Kendall},
  \bibinfo{person}{Shefali Haldar}, \bibinfo{person}{Ari~H. Pollack}, {and}
  \bibinfo{person}{Wanda Pratt}.} \bibinfo{year}{2016}\natexlab{}.
\newblock \showarticletitle{Partners in Care: Design Considerations for
  Caregivers and Patients During a Hospital Stay}. In
  \bibinfo{booktitle}{\emph{Proceedings of the 19th ACM Conference on
  Computer-Supported Cooperative Work \& Social Computing}} (San Francisco,
  California, USA) \emph{(\bibinfo{series}{CSCW '16})}.
  \bibinfo{publisher}{Association for Computing Machinery},
  \bibinfo{address}{New York, NY, USA}, \bibinfo{pages}{756–769}.
\newblock
\showISBNx{9781450335928}
\urldef\tempurl%
\url{https://doi.org/10.1145/2818048.2819983}
\showDOI{\tempurl}


\bibitem[Montagna et~al\mbox{.}(2023)]%
        {montagna2023data}
\bibfield{author}{\bibinfo{person}{Sara Montagna}, \bibinfo{person}{Stefano
  Ferretti}, \bibinfo{person}{Lorenz~Cuno Klopfenstein},
  \bibinfo{person}{Antonio Florio}, {and} \bibinfo{person}{Martino~Francesco
  Pengo}.} \bibinfo{year}{2023}\natexlab{}.
\newblock \showarticletitle{Data decentralisation of LLM-based chatbot systems
  in chronic disease self-management}. In \bibinfo{booktitle}{\emph{Proceedings
  of the 2023 ACM Conference on Information Technology for Social Good}}.
  \bibinfo{pages}{205--212}.
\newblock


\bibitem[Nelson et~al\mbox{.}(2006)]%
        {nelson2006improving}
\bibfield{author}{\bibinfo{person}{JE Nelson}, \bibinfo{person}{CM Mulkerin},
  \bibinfo{person}{LL Adams}, {and} \bibinfo{person}{PJ Pronovost}.}
  \bibinfo{year}{2006}\natexlab{}.
\newblock \showarticletitle{Improving comfort and communication in the ICU: a
  practical new tool for palliative care performance measurement and feedback}.
\newblock \bibinfo{journal}{\emph{BMJ Quality \& Safety}} \bibinfo{volume}{15},
  \bibinfo{number}{4} (\bibinfo{year}{2006}), \bibinfo{pages}{264--271}.
\newblock


\bibitem[Nikkhah et~al\mbox{.}(2021)]%
        {Nikkhah2021}
\bibfield{author}{\bibinfo{person}{Sarah Nikkhah}, \bibinfo{person}{Swaroop
  John}, \bibinfo{person}{Krishna~Supradeep Yalamarti}, \bibinfo{person}{Emily
  L.~Mueller}, {and} \bibinfo{person}{Andrew D.~Miller}.}
  \bibinfo{year}{2021}\natexlab{}.
\newblock \showarticletitle{Helping Their Child, Helping Each Other: Parents’
  Mediated Social Support in the Children's Hospital}. In
  \bibinfo{booktitle}{\emph{Companion Publication of the 2021 Conference on
  Computer Supported Cooperative Work and Social Computing}} (Virtual Event,
  USA) \emph{(\bibinfo{series}{CSCW '21 Companion})}.
  \bibinfo{publisher}{Association for Computing Machinery},
  \bibinfo{address}{New York, NY, USA}, \bibinfo{pages}{140–143}.
\newblock
\showISBNx{9781450384797}
\urldef\tempurl%
\url{https://doi.org/10.1145/3462204.3481759}
\showDOI{\tempurl}


\bibitem[Nyapathy and Arriaga(2019)]%
        {Nyapathy2019}
\bibfield{author}{\bibinfo{person}{Nikhila Nyapathy} {and}
  \bibinfo{person}{Rosa~I. Arriaga}.} \bibinfo{year}{2019}\natexlab{}.
\newblock \showarticletitle{Tracking and Reporting Asthma Data for Children}.
  In \bibinfo{booktitle}{\emph{Companion Publication of the 2019 Conference on
  Computer Supported Cooperative Work and Social Computing}} (Austin, TX, USA)
  \emph{(\bibinfo{series}{CSCW '19 Companion})}.
  \bibinfo{publisher}{Association for Computing Machinery},
  \bibinfo{address}{New York, NY, USA}, \bibinfo{pages}{330–334}.
\newblock
\showISBNx{9781450366922}
\urldef\tempurl%
\url{https://doi.org/10.1145/3311957.3359480}
\showDOI{\tempurl}


\bibitem[Paul et~al\mbox{.}(2004)]%
        {paul2004meeting}
\bibfield{author}{\bibinfo{person}{Fiona Paul}, \bibinfo{person}{Charles
  Hendry}, {and} \bibinfo{person}{Louise Cabrelli}.}
  \bibinfo{year}{2004}\natexlab{}.
\newblock \showarticletitle{Meeting patient and relatives’ information needs
  upon transfer from an intensive care unit: the development and evaluation of
  an information booklet}.
\newblock \bibinfo{journal}{\emph{Journal of clinical nursing}}
  \bibinfo{volume}{13}, \bibinfo{number}{3} (\bibinfo{year}{2004}),
  \bibinfo{pages}{396--405}.
\newblock


\bibitem[Pina et~al\mbox{.}(2017)]%
        {Pina2017}
\bibfield{author}{\bibinfo{person}{Laura~R. Pina}, \bibinfo{person}{Sang-Wha
  Sien}, \bibinfo{person}{Teresa Ward}, \bibinfo{person}{Jason~C. Yip},
  \bibinfo{person}{Sean~A. Munson}, \bibinfo{person}{James Fogarty}, {and}
  \bibinfo{person}{Julie~A. Kientz}.} \bibinfo{year}{2017}\natexlab{}.
\newblock \showarticletitle{From Personal Informatics to Family Informatics:
  Understanding Family Practices around Health Monitoring}. In
  \bibinfo{booktitle}{\emph{Proceedings of the 2017 ACM Conference on Computer
  Supported Cooperative Work and Social Computing}} (Portland, Oregon, USA)
  \emph{(\bibinfo{series}{CSCW '17})}. \bibinfo{publisher}{Association for
  Computing Machinery}, \bibinfo{address}{New York, NY, USA},
  \bibinfo{pages}{2300–2315}.
\newblock
\showISBNx{9781450343350}
\urldef\tempurl%
\url{https://doi.org/10.1145/2998181.2998362}
\showDOI{\tempurl}


\bibitem[Pine et~al\mbox{.}(2018)]%
        {Pine2018}
\bibfield{author}{\bibinfo{person}{Kathleen~H. Pine}, \bibinfo{person}{Claus
  Bossen}, \bibinfo{person}{Yunan Chen}, \bibinfo{person}{Gunnar Ellingsen},
  \bibinfo{person}{Miria Grisot}, \bibinfo{person}{Melissa Mazmanian}, {and}
  \bibinfo{person}{Naja~Holten M\o{}ller}.} \bibinfo{year}{2018}\natexlab{}.
\newblock \showarticletitle{Data Work in Healthcare: Challenges for Patients,
  Clinicians and Administrators}. In \bibinfo{booktitle}{\emph{Companion of the
  2018 ACM Conference on Computer Supported Cooperative Work and Social
  Computing}} (Jersey City, NJ, USA) \emph{(\bibinfo{series}{CSCW '18
  Companion})}. \bibinfo{publisher}{Association for Computing Machinery},
  \bibinfo{address}{New York, NY, USA}, \bibinfo{pages}{433–439}.
\newblock
\showISBNx{9781450360180}
\urldef\tempurl%
\url{https://doi.org/10.1145/3272973.3273017}
\showDOI{\tempurl}


\bibitem[Ploderer et~al\mbox{.}(2016)]%
        {ploderer2016armsleeve}
\bibfield{author}{\bibinfo{person}{Bernd Ploderer}, \bibinfo{person}{Justin
  Fong}, \bibinfo{person}{Anusha Withana}, \bibinfo{person}{Marlena Klaic},
  \bibinfo{person}{Siddharth Nair}, \bibinfo{person}{Vincent Crocher},
  \bibinfo{person}{Frank Vetere}, {and} \bibinfo{person}{Suranga Nanayakkara}.}
  \bibinfo{year}{2016}\natexlab{}.
\newblock \showarticletitle{ArmSleeve: a patient monitoring system to support
  occupational therapists in stroke rehabilitation}. In
  \bibinfo{booktitle}{\emph{Proceedings of the 2016 ACM Conference on Designing
  Interactive Systems}}. \bibinfo{pages}{700--711}.
\newblock


\bibitem[Pronovost et~al\mbox{.}(2003)]%
        {pronovost2003improving}
\bibfield{author}{\bibinfo{person}{Peter Pronovost}, \bibinfo{person}{Sean
  Berenholtz}, \bibinfo{person}{Todd Dorman}, \bibinfo{person}{Pam~A Lipsett},
  \bibinfo{person}{Terri Simmonds}, {and} \bibinfo{person}{Carol Haraden}.}
  \bibinfo{year}{2003}\natexlab{}.
\newblock \showarticletitle{Improving communication in the ICU using daily
  goals}.
\newblock \bibinfo{journal}{\emph{Journal of critical care}}
  \bibinfo{volume}{18}, \bibinfo{number}{2} (\bibinfo{year}{2003}),
  \bibinfo{pages}{71--75}.
\newblock


\bibitem[Ramjee et~al\mbox{.}(2024)]%
        {ramjee2024cataractbot}
\bibfield{author}{\bibinfo{person}{Pragnya Ramjee}, \bibinfo{person}{Bhuvan
  Sachdeva}, \bibinfo{person}{Satvik Golechha}, \bibinfo{person}{Shreyas
  Kulkarni}, \bibinfo{person}{Geeta Fulari}, \bibinfo{person}{Kaushik Murali},
  {and} \bibinfo{person}{Mohit Jain}.} \bibinfo{year}{2024}\natexlab{}.
\newblock \showarticletitle{CataractBot: an LLM-powered expert-in-the-loop
  chatbot for cataract patients}.
\newblock \bibinfo{journal}{\emph{arXiv preprint arXiv:2402.04620}}
  (\bibinfo{year}{2024}).
\newblock


\bibitem[Ramsey et~al\mbox{.}(2018)]%
        {ramsey2018increasing}
\bibfield{author}{\bibinfo{person}{Alexandra Ramsey}, \bibinfo{person}{Erin
  Lanzo}, \bibinfo{person}{Hattie Huston-Paterson}, \bibinfo{person}{Kathy
  Tomaszewski}, {and} \bibinfo{person}{Maria Trent}.}
  \bibinfo{year}{2018}\natexlab{}.
\newblock \showarticletitle{Increasing patient portal usage: preliminary
  outcomes from the MyChart genius project}.
\newblock \bibinfo{journal}{\emph{Journal of Adolescent Health}}
  \bibinfo{volume}{62}, \bibinfo{number}{1} (\bibinfo{year}{2018}),
  \bibinfo{pages}{29--35}.
\newblock


\bibitem[Scheunemann et~al\mbox{.}(2011)]%
        {scheunemann2011randomized}
\bibfield{author}{\bibinfo{person}{Leslie~P Scheunemann},
  \bibinfo{person}{Michelle McDevitt}, \bibinfo{person}{Shannon~S Carson},
  {and} \bibinfo{person}{Laura~C Hanson}.} \bibinfo{year}{2011}\natexlab{}.
\newblock \showarticletitle{Randomized, controlled trials of interventions to
  improve communication in intensive care: a systematic review}.
\newblock \bibinfo{journal}{\emph{Chest}} \bibinfo{volume}{139},
  \bibinfo{number}{3} (\bibinfo{year}{2011}), \bibinfo{pages}{543--554}.
\newblock


\bibitem[Schnock et~al\mbox{.}(2017)]%
        {schnock2017identifying}
\bibfield{author}{\bibinfo{person}{Kumiko~O Schnock},
  \bibinfo{person}{Sucheta~S Ravindran}, \bibinfo{person}{Anne Fladger},
  \bibinfo{person}{Kathleen Leone}, \bibinfo{person}{Donna~M Williams},
  \bibinfo{person}{Cynthia~L Dwyer}, \bibinfo{person}{Thanh-Giang Vu},
  \bibinfo{person}{Kevin Thornton}, {and} \bibinfo{person}{Priscilla
  Gazarian}.} \bibinfo{year}{2017}\natexlab{}.
\newblock \showarticletitle{Identifying information resources for patients in
  the intensive care unit and their families}.
\newblock \bibinfo{journal}{\emph{Critical Care Nurse}} \bibinfo{volume}{37},
  \bibinfo{number}{6} (\bibinfo{year}{2017}), \bibinfo{pages}{e10--e16}.
\newblock


\bibitem[Schnur et~al\mbox{.}(2024)]%
        {Schnur2024}
\bibfield{author}{\bibinfo{person}{Jennifer~J. Schnur},
  \bibinfo{person}{Ang\'{e}lica Garcia-Mart\'{\i}nez}, \bibinfo{person}{Patrick
  Soga}, \bibinfo{person}{Karla Badillo-Urquiola},
  \bibinfo{person}{Alejandra~J. Botello}, \bibinfo{person}{Ana
  Calderon~Raisbeck}, \bibinfo{person}{Sugana Chawla}, \bibinfo{person}{Josef
  Ernst}, \bibinfo{person}{William Gentry}, \bibinfo{person}{Richard~P.
  Johnson}, \bibinfo{person}{Michael Kennel}, \bibinfo{person}{Jes\'{u}s
  Robles}, \bibinfo{person}{Madison Wagner}, \bibinfo{person}{Elizabeth
  Medina}, \bibinfo{person}{Juan Gardu\~{n}o Espinosa},
  \bibinfo{person}{Horacio M\'{a}rquez-Gonz\'{a}lez}, \bibinfo{person}{Victor
  Olivar-L\'{o}pez}, \bibinfo{person}{Luis~E. Ju\'{a}rez-Villegas},
  \bibinfo{person}{Martha Avil\'{e}s-Robles}, \bibinfo{person}{Elisa
  Dorantes-Acosta}, \bibinfo{person}{Viridia Avila}, \bibinfo{person}{Gina
  Chapa-Koloffon}, \bibinfo{person}{Elizabeth Cruz}, \bibinfo{person}{Leticia
  Luis}, \bibinfo{person}{Clara Quezada}, \bibinfo{person}{Emanuel Orozco},
  \bibinfo{person}{Edson Serv\'{a}n-Mori}, \bibinfo{person}{Martha Cordero},
  \bibinfo{person}{Rub\'{e}n Mart\'{\i}n~Payo}, {and}
  \bibinfo{person}{Nitesh~V. Chawla}.} \bibinfo{year}{2024}\natexlab{}.
\newblock \showarticletitle{SaludConectaMX: Lessons Learned from Deploying a
  Cooperative Mobile Health System for Pediatric Cancer Care in Mexico}. In
  \bibinfo{booktitle}{\emph{Companion Publication of the 2024 Conference on
  Computer-Supported Cooperative Work and Social Computing}} (San Jose, Costa
  Rica) \emph{(\bibinfo{series}{CSCW Companion '24})}.
  \bibinfo{publisher}{Association for Computing Machinery},
  \bibinfo{address}{New York, NY, USA}, \bibinfo{pages}{316–322}.
\newblock
\showISBNx{9798400711145}
\urldef\tempurl%
\url{https://doi.org/10.1145/3678884.3685922}
\showDOI{\tempurl}


\bibitem[Shea et~al\mbox{.}(2019)]%
        {Shea2019}
\bibfield{author}{\bibinfo{person}{Zachary Shea}, \bibinfo{person}{Zaina
  Aljallad}, \bibinfo{person}{David Taylor}, \bibinfo{person}{Chhaya Chouhan},
  {and} \bibinfo{person}{Pamela~J. Wisniewski}.}
  \bibinfo{year}{2019}\natexlab{}.
\newblock \showarticletitle{Carebit: A Mobile App for Remote Informal
  Caregiving}. In \bibinfo{booktitle}{\emph{Companion Publication of the 2019
  Conference on Computer Supported Cooperative Work and Social Computing}}
  (Austin, TX, USA) \emph{(\bibinfo{series}{CSCW '19 Companion})}.
  \bibinfo{publisher}{Association for Computing Machinery},
  \bibinfo{address}{New York, NY, USA}, \bibinfo{pages}{23–27}.
\newblock
\showISBNx{9781450366922}
\urldef\tempurl%
\url{https://doi.org/10.1145/3311957.3359511}
\showDOI{\tempurl}


\bibitem[Siddiqui et~al\mbox{.}(2023)]%
        {Siddiqui2023}
\bibfield{author}{\bibinfo{person}{Farheen Siddiqui}, \bibinfo{person}{Delvin
  Varghese}, \bibinfo{person}{Pushpendra Singh}, \bibinfo{person}{Sunita~Bapuji
  Bayyavarapu}, \bibinfo{person}{Stephen Lindsay}, \bibinfo{person}{Dharshani
  Chandrasekara}, \bibinfo{person}{Pranav Kulkarni}, \bibinfo{person}{Ling Wu},
  \bibinfo{person}{Taghreed Alshehri}, {and} \bibinfo{person}{Patrick
  Olivier}.} \bibinfo{year}{2023}\natexlab{}.
\newblock \showarticletitle{Exploring the digital support needs of caregivers
  of people with serious mental illness}. In
  \bibinfo{booktitle}{\emph{Proceedings of the 2023 CHI Conference on Human
  Factors in Computing Systems}} (Hamburg, Germany) \emph{(\bibinfo{series}{CHI
  '23})}. \bibinfo{publisher}{Association for Computing Machinery},
  \bibinfo{address}{New York, NY, USA}, Article \bibinfo{articleno}{560},
  \bibinfo{numpages}{16}~pages.
\newblock
\showISBNx{9781450394215}
\urldef\tempurl%
\url{https://doi.org/10.1145/3544548.3580674}
\showDOI{\tempurl}


\bibitem[Smriti et~al\mbox{.}(2024)]%
        {smriti2024emotion}
\bibfield{author}{\bibinfo{person}{Diva Smriti}, \bibinfo{person}{Lu Wang},
  {and} \bibinfo{person}{Jina Huh-Yoo}.} \bibinfo{year}{2024}\natexlab{}.
\newblock \showarticletitle{Emotion work in caregiving: the role of technology
  to support informal caregivers of persons living with dementia}.
\newblock \bibinfo{journal}{\emph{Proceedings of the ACM on human-computer
  interaction}} \bibinfo{volume}{8}, \bibinfo{number}{CSCW1}
  (\bibinfo{year}{2024}), \bibinfo{pages}{1--34}.
\newblock


\bibitem[Steel et~al\mbox{.}(2008)]%
        {steel2008impact}
\bibfield{author}{\bibinfo{person}{Alistair Steel}, \bibinfo{person}{Carol
  Underwood}, \bibinfo{person}{Caitlin Notley}, {and} \bibinfo{person}{Mark
  Blunt}.} \bibinfo{year}{2008}\natexlab{}.
\newblock \showarticletitle{The impact of offering a relatives’ clinic on the
  satisfaction of the next-of-kin of critical care patients—A prospective
  time-interrupted trial}.
\newblock \bibinfo{journal}{\emph{Intensive and Critical Care Nursing}}
  \bibinfo{volume}{24}, \bibinfo{number}{2} (\bibinfo{year}{2008}),
  \bibinfo{pages}{122--129}.
\newblock


\bibitem[Stefanidi et~al\mbox{.}(2023)]%
        {Stefanidi2023}
\bibfield{author}{\bibinfo{person}{Evropi Stefanidi}, \bibinfo{person}{Johannes
  Sch\"{o}ning}, \bibinfo{person}{Yvonne Rogers}, {and} \bibinfo{person}{Jasmin
  Niess}.} \bibinfo{year}{2023}\natexlab{}.
\newblock \showarticletitle{Children with ADHD and their Care Ecosystem:
  Designing Beyond Symptoms}. In \bibinfo{booktitle}{\emph{Proceedings of the
  2023 CHI Conference on Human Factors in Computing Systems}} (Hamburg,
  Germany) \emph{(\bibinfo{series}{CHI '23})}. \bibinfo{publisher}{Association
  for Computing Machinery}, \bibinfo{address}{New York, NY, USA}, Article
  \bibinfo{articleno}{558}, \bibinfo{numpages}{17}~pages.
\newblock
\showISBNx{9781450394215}
\urldef\tempurl%
\url{https://doi.org/10.1145/3544548.3581216}
\showDOI{\tempurl}


\bibitem[Suresh et~al\mbox{.}(2017)]%
        {suresh2017clinical}
\bibfield{author}{\bibinfo{person}{Harini Suresh}, \bibinfo{person}{Nathan
  Hunt}, \bibinfo{person}{Alistair Johnson}, \bibinfo{person}{Leo~Anthony
  Celi}, \bibinfo{person}{Peter Szolovits}, {and} \bibinfo{person}{Marzyeh
  Ghassemi}.} \bibinfo{year}{2017}\natexlab{}.
\newblock \showarticletitle{Clinical intervention prediction and understanding
  using deep networks}.
\newblock \bibinfo{journal}{\emph{arXiv preprint arXiv:1705.08498}}
  (\bibinfo{year}{2017}).
\newblock


\bibitem[Tabah et~al\mbox{.}(2022)]%
        {tabah2022variation}
\bibfield{author}{\bibinfo{person}{Alexis Tabah}, \bibinfo{person}{Muhammed
  Elhadi}, \bibinfo{person}{Emma Ballard}, \bibinfo{person}{Andrea Cortegiani},
  \bibinfo{person}{Maurizio Cecconi}, \bibinfo{person}{Takeshi Unoki},
  \bibinfo{person}{Laur{\k{a}} Galarza}, \bibinfo{person}{Regis~Goulart Rosa},
  \bibinfo{person}{Francois Barbier}, \bibinfo{person}{Elie Azoulay},
  {et~al\mbox{.}}} \bibinfo{year}{2022}\natexlab{}.
\newblock \showarticletitle{Variation in communication and family visiting
  policies in intensive care within and between countries during the Covid-19
  pandemic: the COVISIT international survey}.
\newblock \bibinfo{journal}{\emph{Journal of critical care}}
  \bibinfo{volume}{71} (\bibinfo{year}{2022}), \bibinfo{pages}{154050}.
\newblock


\bibitem[Thomas(2006)]%
        {thomas2006general}
\bibfield{author}{\bibinfo{person}{David~R Thomas}.}
  \bibinfo{year}{2006}\natexlab{}.
\newblock \showarticletitle{A general inductive approach for analyzing
  qualitative evaluation data}.
\newblock \bibinfo{journal}{\emph{American journal of evaluation}}
  \bibinfo{volume}{27}, \bibinfo{number}{2} (\bibinfo{year}{2006}),
  \bibinfo{pages}{237--246}.
\newblock


\bibitem[Valentin et~al\mbox{.}(2011)]%
        {valentin2011recommendations}
\bibfield{author}{\bibinfo{person}{Andreas Valentin}, \bibinfo{person}{Patrick
  Ferdinande}, {and} \bibinfo{person}{ESICM Working~Group on
  Quality~Improvement}.} \bibinfo{year}{2011}\natexlab{}.
\newblock \showarticletitle{Recommendations on basic requirements for intensive
  care units: structural and organizational aspects}.
\newblock \bibinfo{journal}{\emph{Intensive care medicine}}
  \bibinfo{volume}{37} (\bibinfo{year}{2011}), \bibinfo{pages}{1575--1587}.
\newblock


\bibitem[Wachter and Brynjolfsson(2024)]%
        {wachter2024will}
\bibfield{author}{\bibinfo{person}{Robert~M Wachter} {and}
  \bibinfo{person}{Erik Brynjolfsson}.} \bibinfo{year}{2024}\natexlab{}.
\newblock \showarticletitle{Will generative artificial intelligence deliver on
  its promise in health care?}
\newblock \bibinfo{journal}{\emph{Jama}} \bibinfo{volume}{331},
  \bibinfo{number}{1} (\bibinfo{year}{2024}), \bibinfo{pages}{65--69}.
\newblock


\bibitem[Wang et~al\mbox{.}(2023)]%
        {wang2023}
\bibfield{author}{\bibinfo{person}{Peng-Jui Wang}, \bibinfo{person}{Yi-Chi
  Lee}, \bibinfo{person}{Uei-Dar Chen}, {and} \bibinfo{person}{Yung-Ju Chang}.}
  \bibinfo{year}{2023}\natexlab{}.
\newblock \showarticletitle{NotiSummary: Exploring the Potential of AI-Driven
  Text Summarization on Smartphone Notification Management}. In
  \bibinfo{booktitle}{\emph{Adjunct Proceedings of the 2023 ACM International
  Joint Conference on Pervasive and Ubiquitous Computing \& the 2023 ACM
  International Symposium on Wearable Computing}} (Cancun, Quintana Roo,
  Mexico) \emph{(\bibinfo{series}{UbiComp/ISWC '23 Adjunct})}.
  \bibinfo{publisher}{Association for Computing Machinery},
  \bibinfo{address}{New York, NY, USA}, \bibinfo{pages}{113–117}.
\newblock
\showISBNx{9798400702006}
\urldef\tempurl%
\url{https://doi.org/10.1145/3594739.3610702}
\showDOI{\tempurl}


\bibitem[Wei et~al\mbox{.}(2024)]%
        {wei2024leveraging}
\bibfield{author}{\bibinfo{person}{Jing Wei}, \bibinfo{person}{Sungdong Kim},
  \bibinfo{person}{Hyunhoon Jung}, {and} \bibinfo{person}{Young-Ho Kim}.}
  \bibinfo{year}{2024}\natexlab{}.
\newblock \showarticletitle{Leveraging large language models to power chatbots
  for collecting user self-reported data}.
\newblock \bibinfo{journal}{\emph{Proceedings of the ACM on Human-Computer
  Interaction}} \bibinfo{volume}{8}, \bibinfo{number}{CSCW1}
  (\bibinfo{year}{2024}), \bibinfo{pages}{1--35}.
\newblock


\bibitem[Wei et~al\mbox{.}(2022)]%
        {wei2022chain}
\bibfield{author}{\bibinfo{person}{Jason Wei}, \bibinfo{person}{Xuezhi Wang},
  \bibinfo{person}{Dale Schuurmans}, \bibinfo{person}{Maarten Bosma},
  \bibinfo{person}{Fei Xia}, \bibinfo{person}{Ed Chi}, \bibinfo{person}{Quoc~V
  Le}, \bibinfo{person}{Denny Zhou}, {et~al\mbox{.}}}
  \bibinfo{year}{2022}\natexlab{}.
\newblock \showarticletitle{Chain-of-thought prompting elicits reasoning in
  large language models}.
\newblock \bibinfo{journal}{\emph{Advances in neural information processing
  systems}}  \bibinfo{volume}{35} (\bibinfo{year}{2022}),
  \bibinfo{pages}{24824--24837}.
\newblock


\bibitem[Withanage~Don et~al\mbox{.}(2024)]%
        {Don2024}
\bibfield{author}{\bibinfo{person}{Daksitha~Senel Withanage~Don},
  \bibinfo{person}{Dominik Schiller}, \bibinfo{person}{Tobias Hallmen},
  \bibinfo{person}{Silvan Mertes}, \bibinfo{person}{Tobias Baur},
  \bibinfo{person}{Florian Lingenfelser}, \bibinfo{person}{Mitho M\"{u}ller},
  \bibinfo{person}{Lea Kaubisch}, \bibinfo{person}{Prof. Dr.~Corinna Reck},
  {and} \bibinfo{person}{Elisabeth Andr\'{e}}.}
  \bibinfo{year}{2024}\natexlab{}.
\newblock \showarticletitle{Towards Automated Annotation of Infant-Caregiver
  Engagement Phases with Multimodal Foundation Models}. In
  \bibinfo{booktitle}{\emph{Proceedings of the 26th International Conference on
  Multimodal Interaction}} (San Jose, Costa Rica) \emph{(\bibinfo{series}{ICMI
  '24})}. \bibinfo{publisher}{Association for Computing Machinery},
  \bibinfo{address}{New York, NY, USA}, \bibinfo{pages}{428–438}.
\newblock
\showISBNx{9798400704628}
\urldef\tempurl%
\url{https://doi.org/10.1145/3678957.3685704}
\showDOI{\tempurl}


\bibitem[Wong et~al\mbox{.}(2018)]%
        {wong2018using}
\bibfield{author}{\bibinfo{person}{Jenna Wong}, \bibinfo{person}{Mara
  Murray~Horwitz}, \bibinfo{person}{Li Zhou}, {and} \bibinfo{person}{Sengwee
  Toh}.} \bibinfo{year}{2018}\natexlab{}.
\newblock \showarticletitle{Using machine learning to identify health outcomes
  from electronic health record data}.
\newblock \bibinfo{journal}{\emph{Current epidemiology reports}}
  \bibinfo{volume}{5} (\bibinfo{year}{2018}), \bibinfo{pages}{331--342}.
\newblock


\bibitem[Wu et~al\mbox{.}(2024)]%
        {wu2024clinical}
\bibfield{author}{\bibinfo{person}{Siyi Wu}, \bibinfo{person}{Weidan Cao},
  \bibinfo{person}{Shihan Fu}, \bibinfo{person}{Bingsheng Yao},
  \bibinfo{person}{Ziqi Yang}, \bibinfo{person}{Changchang Yin},
  \bibinfo{person}{Varun Mishra}, \bibinfo{person}{Daniel Addison},
  \bibinfo{person}{Ping Zhang}, {and} \bibinfo{person}{Dakuo Wang}.}
  \bibinfo{year}{2024}\natexlab{}.
\newblock \showarticletitle{Clinical Challenges and AI Opportunities in
  Decision-Making for Cancer Treatment-Induced Cardiotoxicity}.
\newblock \bibinfo{journal}{\emph{arXiv preprint arXiv:2408.03586}}
  (\bibinfo{year}{2024}).
\newblock


\bibitem[Yang et~al\mbox{.}(2024)]%
        {ziqi2024}
\bibfield{author}{\bibinfo{person}{Ziqi Yang}, \bibinfo{person}{Xuhai Xu},
  \bibinfo{person}{Bingsheng Yao}, \bibinfo{person}{Ethan Rogers},
  \bibinfo{person}{Shao Zhang}, \bibinfo{person}{Stephen Intille},
  \bibinfo{person}{Nawar Shara}, \bibinfo{person}{Guodong~Gordon Gao}, {and}
  \bibinfo{person}{Dakuo Wang}.} \bibinfo{year}{2024}\natexlab{}.
\newblock \showarticletitle{Talk2Care: An LLM-based Voice Assistant for
  Communication between Healthcare Providers and Older Adults}.
\newblock \bibinfo{journal}{\emph{Proc. ACM Interact. Mob. Wearable Ubiquitous
  Technol.}} \bibinfo{volume}{8}, \bibinfo{number}{2}, Article
  \bibinfo{articleno}{73} (\bibinfo{date}{May} \bibinfo{year}{2024}),
  \bibinfo{numpages}{35}~pages.
\newblock
\urldef\tempurl%
\url{https://doi.org/10.1145/3659625}
\showDOI{\tempurl}


\bibitem[Yim et~al\mbox{.}(2024)]%
        {yim2024preliminary}
\bibfield{author}{\bibinfo{person}{Dobin Yim}, \bibinfo{person}{Jiban Khuntia},
  \bibinfo{person}{Vijaya Parameswaran}, \bibinfo{person}{Arlen Meyers},
  {et~al\mbox{.}}} \bibinfo{year}{2024}\natexlab{}.
\newblock \showarticletitle{Preliminary Evidence of the Use of Generative AI in
  Health Care Clinical Services: Systematic Narrative Review}.
\newblock \bibinfo{journal}{\emph{JMIR Medical Informatics}}
  \bibinfo{volume}{12}, \bibinfo{number}{1} (\bibinfo{year}{2024}),
  \bibinfo{pages}{e52073}.
\newblock


\bibitem[Yin et~al\mbox{.}(2024)]%
        {yin2024sepsiscalc}
\bibfield{author}{\bibinfo{person}{Changchang Yin}, \bibinfo{person}{Shihan
  Fu}, \bibinfo{person}{Bingsheng Yao}, \bibinfo{person}{Thai-Hoang Pham},
  \bibinfo{person}{Weidan Cao}, \bibinfo{person}{Dakuo Wang},
  \bibinfo{person}{Jeffrey Caterino}, {and} \bibinfo{person}{Ping Zhang}.}
  \bibinfo{year}{2024}\natexlab{}.
\newblock \showarticletitle{SepsisCalc: Integrating Clinical Calculators into
  Early Sepsis Prediction via Dynamic Temporal Graph Construction}.
\newblock \bibinfo{journal}{\emph{arXiv preprint arXiv:2501.00190}}
  (\bibinfo{year}{2024}).
\newblock


\bibitem[Yoo et~al\mbox{.}(2020)]%
        {yoo2020critical}
\bibfield{author}{\bibinfo{person}{Hye~Jin Yoo}, \bibinfo{person}{Oak~Bun Lim},
  {and} \bibinfo{person}{Jae~Lan Shim}.} \bibinfo{year}{2020}\natexlab{}.
\newblock \showarticletitle{Critical care nurses’ communication experiences
  with patients and families in an intensive care unit: A qualitative study}.
\newblock \bibinfo{journal}{\emph{Plos one}} \bibinfo{volume}{15},
  \bibinfo{number}{7} (\bibinfo{year}{2020}), \bibinfo{pages}{e0235694}.
\newblock


\bibitem[Young et~al\mbox{.}(2017)]%
        {young2017family}
\bibfield{author}{\bibinfo{person}{Amanda~J Young}, \bibinfo{person}{Elizabeth
  Stephens}, {and} \bibinfo{person}{Joy~V Goldsmith}.}
  \bibinfo{year}{2017}\natexlab{}.
\newblock \showarticletitle{Family caregiver communication in the ICU: Toward a
  relational view of health literacy}.
\newblock \bibinfo{journal}{\emph{Journal of Family Communication}}
  \bibinfo{volume}{17}, \bibinfo{number}{2} (\bibinfo{year}{2017}),
  \bibinfo{pages}{137--152}.
\newblock


\bibitem[Zhang et~al\mbox{.}(2024)]%
        {zhang2024}
\bibfield{author}{\bibinfo{person}{Shao Zhang}, \bibinfo{person}{Jianing Yu},
  \bibinfo{person}{Xuhai Xu}, \bibinfo{person}{Changchang Yin},
  \bibinfo{person}{Yuxuan Lu}, \bibinfo{person}{Bingsheng Yao},
  \bibinfo{person}{Melanie Tory}, \bibinfo{person}{Lace~M. Padilla},
  \bibinfo{person}{Jeffrey Caterino}, \bibinfo{person}{Ping Zhang}, {and}
  \bibinfo{person}{Dakuo Wang}.} \bibinfo{year}{2024}\natexlab{}.
\newblock \showarticletitle{Rethinking Human-AI Collaboration in Complex
  Medical Decision Making: A Case Study in Sepsis Diagnosis}. In
  \bibinfo{booktitle}{\emph{Proceedings of the 2024 CHI Conference on Human
  Factors in Computing Systems}} (Honolulu, HI, USA)
  \emph{(\bibinfo{series}{CHI '24})}. \bibinfo{publisher}{Association for
  Computing Machinery}, \bibinfo{address}{New York, NY, USA}, Article
  \bibinfo{articleno}{445}, \bibinfo{numpages}{18}~pages.
\newblock
\showISBNx{9798400703300}
\urldef\tempurl%
\url{https://doi.org/10.1145/3613904.3642343}
\showDOI{\tempurl}


\bibitem[Zhang et~al\mbox{.}(2023)]%
        {zhang2023ehr}
\bibfield{author}{\bibinfo{person}{Xiaocheng Zhang}, \bibinfo{person}{Zonghai
  Yao}, {and} \bibinfo{person}{Hong Yu}.} \bibinfo{year}{2023}\natexlab{}.
\newblock \showarticletitle{Ehr interaction between patients and ai: Noteaid
  ehr interaction}.
\newblock \bibinfo{journal}{\emph{arXiv preprint arXiv:2312.17475}}
  (\bibinfo{year}{2023}).
\newblock


\end{thebibliography}

\appendix

\section{Formative Study - Interview Script}\label{Study1script}

\begin{itemize}
    \item \textbf{Question 1 - Experience:} 
    Can you describe your experience as a caregiver when your loved one was transferred to the ICU?

    \item \textbf{Question 2 - Information:}
    What kind of information did you seek as a caregiver and what was your experience accessing and understanding the information?
    
    \item \textbf{Question 3 - Technology Use:} How did you utilize digital tools, such as the patient portal or search engines, to gather information?

    \item \textbf{Question 4 - Technology Expectation:} What do you expect of an AI-based digital tool that is designed to help you in such a situation? Would you trust the AI?


\end{itemize}

\section{Evaluation Study}\label{Study2script}

\subsection{Interview Script}
The following sets of questions were asked for each interface presented to the participants.

\subsubsection*{Family Caregivers}
\begin{itemize}
    \item \textbf{Question 1 – Feedback on the Design:} (After showing a screenshot of the UI) How do you like this interface?

    \item \textbf{Question 2 – Utility:} Do you think this function would be useful to you? Why or why not?

    \item \textbf{Question 3 – Use Cases:} How would you use this interface/function? [What kind of questions would you ask this chatbot?]

    \item \textbf{Question 4 – Improvement:} What improvements can be made?
\end{itemize}

\subsubsection*{Clinical Experts (Physicians)}
\begin{itemize}
    \item \textbf{Question 1 – Output Quality:} How accurate, clear, and clinically appropriate do you find the system-generated outputs (daily summaries, treatment plans, chatbot responses)?

    \item \textbf{Question 2 – Clinical Utility:} Do you think this system could support clinicians or improve communication with caregivers? Why or why not?

    \item \textbf{Question 3 – Usability Feedback:} From a clinical perspective, how usable and trustworthy do you find this interface?

    \item \textbf{Question 4 – Improvement:} What improvements would you suggest to make the system more clinically reliable and usable?
\end{itemize}

\subsection{Questionnaire - Clinician Feedback on AI-generated} \label{questionnaire}
Please rate your level of agreement with the following statements about the AI-generated summary. 
1 - strongly disagree, 5 - strongly agree. 
\begin{itemize}
  \item \textbf{Overall Accuracy} - The AI-generated output presents medically accurate information without major errors.
  \item \textbf{Hallucination} - The AI-generated output does not contain fabricated or factually incorrect information.
  \item \textbf{Readability} - The AI-generated output is written in clear, well-structured language that is easy to read and understand quickly.
  \item \textbf{Specificity} - The AI-generated output provides sufficient specific details (e.g., vital signs, medication names, test results) rather than vague or general statements.
  \item  \textbf{Overall Quality} - Overall, the AI-generated output is clear, trustworthy, and helpful for quickly understanding the patient’s daily status.
\end{itemize}

\section{Qualitative Codebook}
\onecolumn

\begin{table}[h!]
\centering

\caption{Qualitative Codebook of Formative Study Findings (Stage 1)}
\label{Appendix:codebookofformative}
\resizebox{\textwidth}{!}{%
\begin{tabular}{p{0.19\textwidth}|p{0.19\textwidth}|p{0.6\textwidth}}
\toprule
\textbf{Theme} & \textbf{Sub-Theme} & \textbf{Example} \\
\midrule

\textbf{Challenges in Accessing Clinical Information}
& \multirow{3}{=}{Staying Informed about Patients' Changing Status}  
& "It's hard to get to know who's on the shift that day, and sometimes you have to go there physically [to know his situation]." (P4) \\
\cmidrule{3-3}
& & "There were times when I had to seek out doctors or nurses for updates,... I sometimes felt that my patients were seen as an interruption to the busy ICU schedule... In cases where I didn't get information, I didn't get to know what was going on,... I just felt I was left in the dark" (P1) \\
\cmidrule{3-3}
& & "They are also doing the other stuff, [like] attending to other patients, I wouldn't want to like bother them that much... they are too busy you weren't able to ask all the questions you want to, [and] they weren't able to like explain in details within the short time." (P12) \\
\cmidrule{2-3}

& \multirow{3}{=}{Struggling to Align Treatment Plans and Goals with Healthcare Team}
& "So they told me that,... We should probably be more than a week at the ICU and just informing me about... like they might have to maybe perform another surgery if he... doesn't get responsive, and just telling me other treatment options, like ultrafiltration that the end add transplants that might be considered if it isn't responsive." (P6) \\
\cmidrule{3-3}
& & "[Clinicians did] all the tests, the procedure - [but] what they tested and what they saw, what they got, that kind of deep parts - [I don't know]. I wanted to know the whole [situation] about it, but I think they didn't open it all up, and they were just telling us the diagnosis." (P7) \\
\cmidrule{3-3}
& & "I don't feel like [I'm totally involved in the treatment plan]...Most of the time, the treatment plan is actually well explained, but I do feel like the side effects of certain medications [are] not actually being spelled out." (P8) \\

\midrule

\textbf{Challenges in Understanding Clinical Information}
& \multirow{3}{=} {Interpreting Medical terms in Clinical Report and Conversations} 
& "Some doctors tend to use ambiguous terms..." (P4) \\
\cmidrule{3-3}
& & "We couldn't understand some medical terms. So we asked him (the doctor) to use a clear word where he explained everything that is happening around..." (P5) \\
\cmidrule{3-3}
& & "when it comes to having to pick reports,...the report is not actually something I could compliment, because there are times there that I do have to... use Google to actually go through and know what it actually means." (P8) \\
\cmidrule{2-3}
& \multirow{2}{=}{Asking meaningful questions with healthcare team}
& "Because there was this time I did ask the question, and I was told, even if I did explain to you, you wouldn't understand. Just believe that we are doing this for the good of your relative. You see, so for someone who's afraid to know what they should know, [their choice] is not going to ask questions, which is very wrong." (P4) \\
\cmidrule{3-3}

& & "I was on myself, so I do ask random questions [like] will it be fine? How much is it going to cost us to be here? How long are we going to be here?" (P11) \\

\bottomrule
\end{tabular}
}
\end{table}

\begin{table}[h!]
\centering

\caption{Qualitative Codebook of Prototype Evaluation Study Findings (Stage 2)}
\label{Appendix:codebookofforevaluation}
\resizebox{\textwidth}{!}{%
\begin{tabular}{p{0.19\textwidth}|p{0.19\textwidth}|p{0.6\textwidth}}
\toprule
\textbf{Theme} & \textbf{Sub-Theme} & \textbf{Example} \\
\midrule

\textbf{Visual Timeline of ICU Stay: Supporting Caregivers’ Access to Consistent Information}
& \multirow{3}{=}{Synthesizing Fragmented Information into a Cohesive Timeline}  
& "[The medical timeline] should be helpful to monitor the progress, and if [the patient's condition is] deteriorating or [if] the treatment is being progressive" (P6) \\
\cmidrule{3-3}
& & "[I] might not be daily [to check on that information], but anytime you want to really know about what is really wrong, to have more understanding [on the patient's status], you just look at it." (P7) \\
\cmidrule{2-3}

& \multirow{3}{=}{Balancing Simplicity and Depth in AI-Generated Summaries}
& "Any [other] information I would like to see is the previous illness of the patient... if the patient is having any other illness [that leads to this result]." (P10) \\
\cmidrule{3-3}
& & "[I want to know more details about] after the test is conducted, either a blood test or any test, what they found out after tests." (P7) \\

\cmidrule{2-3}

& \multirow{3}{=}{Involving Caregivers in Treatment Plans and Goals}
& "`Yeah. It[the goals] like the kind of results, [and improvements] that we should be expecting, that's the only thing I noticed" (P9) \\
\cmidrule{3-3}
& & "[besides providing the treatment, I want to know] how he is responding to treatment, [based on] the kind of treatment has been given that day, and the activities of the day." (P10) \\

\midrule

\textbf{Context-Aware LLM-Based Chatbot: Enhancing Caregivers’ Understanding of ICU Information}
& \multirow{3}{=} {Providing Reliable, On-Demand Responses} 
& "It helps because there are times ... instead of waiting to see a doctor to actually explain [the treatment plan], you could actually just use the [chatbot] in this portal to know the reason why a particular equipment or a particular thing is actually administered." (P8) \\
\cmidrule{3-3}
& & "I wouldn't like one hundred percent trust the chatbot that answers my question. But I think with time, I might verify [the] answers [by] speaking to a doctor." (P15) \\
\cmidrule{2-3}
& \multirow{2}{=}{Translating Complex Data into Clear Insights}
& "[I would ask] what some references mean and what they could mean for the patient." (P13) \\
\cmidrule{3-3}

& & "I think having past diagnosis [connected to this chatbot] would be helpful... [then] you know [the] diagnose [is] proper, [and] without having any side effects [to the patients]." (P15) \\

\cmidrule{2-3}
& \multirow{2}{=}{Structuring Conversations and Improving Health Literacy}
& "``Instead of racking my brain on what to ask the doctor,... I'll just like go straight to the point instead of trying to find things out, I'll just ask doctors the possible questions that I got from [the chatbot]." (P12) \\
\cmidrule{3-3}

& & "Just the questions [listed] here would help me prepare questions [for the conversations], because the doctor mentioned so many things that I wouldn't even recall [and] I wouldn't even understand. But if I had prepared my question before and after [that] the doctor answered, then it helped me understand the situation" (P6) \\

\bottomrule
\end{tabular}
}
\end{table}

\end{document}